\documentclass[journal]{IEEEtran}
\ifCLASSINFOpdf
   \usepackage[pdftex]{graphicx}
   \graphicspath{{../pdf/}{../jpeg/}}
   \DeclareGraphicsExtensions{.pdf,.jpeg,.png}
\else
   \usepackage[dvips]{graphicx}
   \graphicspath{{../eps/}}
   \DeclareGraphicsExtensions{.eps}
\fi
\usepackage{epstopdf}
\usepackage{bm}
\usepackage{amsthm}
\UseRawInputEncoding
\usepackage{subfigure}
\usepackage{caption}
\usepackage{cite}
\usepackage{color}
\usepackage{amsmath,amssymb,amsfonts}
\usepackage{textcomp}
\usepackage{xcolor}
\usepackage{array}
\usepackage{booktabs}
\usepackage{multirow}
\usepackage{url}
\usepackage{stfloats}
\def\BibTeX{{\rm B\kern-.05em{\sc i\kern-.025em b}\kern-.08em
    T\kern-.1667em\lower.7ex\hbox{E}\kern-.125emX}}
\begin{document}

\title{On Secrecy Performance of RIS-Assisted MISO Systems over Rician Channels with\\ Spatially Random Eavesdroppers}

\author{
Wei~Shi,~\IEEEmembership{Graduate Student Member,~IEEE}, Jindan~Xu,~\IEEEmembership{Member,~IEEE}, Wei~Xu,~\IEEEmembership{Senior Member,~IEEE}, Chau~Yuen,~\IEEEmembership{Fellow,~IEEE}, A.~Lee~Swindlehurst,~\IEEEmembership{Fellow,~IEEE}, and Chunming~Zhao,~\IEEEmembership{Member,~IEEE} \vspace{-15.pt}

\thanks{W. Shi, W. Xu, and C. Zhao are with the National Mobile Communications Research Laboratory, Southeast University, Nanjing 210096, China, and are also with the Purple Mountain Laboratories, Nanjing 211111, China (e-mail: \{wshi, wxu, cmzhao\}@seu.edu.cn).}

\thanks{J. Xu and C. Yuen are with the School of Electrical and Electronics Engineering, Nanyang Technological University, Singapore 639798, Singapore (e-mail: jindan1025@gmail.com, chau.yuen@ntu.edu.sg).}

\thanks{A. L. Swindlehurst is with the Center for Pervasive Communications and Computing, Henry Samueli School of Engineering, University of California at Irvine, Irvine, CA 92697 USA (e-mail: swindle@uci.edu).}

\thanks{Part of this paper was accepted by IEEE GLOBECOM 2023 Workshops (GC Wkshps) \cite{400}.}}

\maketitle

\begin{abstract}

Reconfigurable intelligent surface (RIS) technology is emerging as a promising technique for performance enhancement for next-generation wireless networks. This paper investigates the physical layer security of an RIS-assisted multiple-antenna communication system in the presence of random spatially distributed eavesdroppers. The RIS-to-ground channels are assumed to experience Rician fading. Using stochastic geometry, exact distributions of the received signal-to-noise-ratios (SNRs) at the legitimate user and the eavesdroppers located according to a Poisson point process (PPP) are derived, and closed-form expressions for the secrecy outage probability (SOP) and the ergodic secrecy capacity (ESC) are obtained to provide insightful guidelines for system design. 
First, the secrecy diversity order is obtained as $\frac{2}{\alpha_2}$, where $\alpha_2$ denotes the path loss exponent of the RIS-to-ground links.
Then, it is revealed that the secrecy performance is mainly affected by the number of RIS reflecting elements, $N$, and the impact of the number of transmit antennas and transmit power at the base station is marginal.
In addition, when the locations of the randomly located eavesdroppers are unknown, deploying the RIS closer to the legitimate user rather than to the base station is shown to be more efficient.
Moreover, it is also found that the density of randomly located eavesdroppers, $\lambda_e$, has an additive effect on the asymptotic ESC performance given by $\log_2{\left({1}/{\lambda_e}\right)}$.
Finally, numerical simulations are conducted to verify the accuracy of these theoretical observations.


\end{abstract}

\section{Introduction}
Reconfigurable intelligent surface (RIS) technology has recently been recognized as a promising approach for realizing both spectral and energy efficient communications in future wireless networks \cite{1,2,200,101}. An RIS comprises a large number of low-cost passive reflecting elements that are able to independently control the phase shifts and/or amplitudes of their reflection coefficients. In this way, the RIS can realize accurate beamforming for adjusting the propagation environments and thus improving the signal quality at desired receivers. In addition, unlike traditional relay transmission, an RIS with miniaturized circuits does not generate new signals or thermal noise. Hence, RISs can be flexibly installed on outdoor buildings, signage, street lamps, and indoor ceilings to help provide additional high-quality links \cite{2}. Due to these advantages, RISs have been widely studied to support a broad range of communication requirements, including data rate enhancement \cite{102,103,121}, coverage extension \cite{104,105,106}, and interference mitigation \cite{107,107-2,122,119}.

In recent years, with the fast-growing number of wireless devices, security for wireless communication has become a critical issue. As a complement to conventional complicated cryptographic methods, physical layer security (PLS) approach leverages the physical characteristics of the propagation environment for enhancing cellular network security against eavesdropping attacks. The capability of an RIS to create a smart controllable wireless propagation environment makes it a promising approach for providing PLS \cite{120}. For example, the authors of \cite{108} and \cite{109} investigated the joint optimization of the active and nearly passive beamforming at the transmitter and the RIS to maximize the theoretical secrecy rate. Furthermore, the design of artificial noise (AN) was also considered in \cite{110} for maximizing the system sum-rate while limiting information leakage to potential eavesdroppers.

On the other hand, there are multiple works that investigate the theoretical secrecy performance for RIS-enhanced PLS systems in terms of secrecy outage probability (SOP) and ergodic secrecy capacity (ESC) \cite{111,3,112,113,114,1003,1004}. In particular, the SOP of an RIS-aided single-antenna system was first studied in the presence of an eavesdropper \cite{111}. An SOP and ESC analysis that considered implementation issues was conducted in \cite{3} and \cite{112}, respectively, under the assumption of discrete RIS phase shifts. RIS-aided secure communications were also studied in emerging applications, such as Device-to-Device (D2D) \cite{113}, vehicular networks \cite{114}, unmanned aerial vehicle (UAV) \cite{1003}, and non-terrestrial networks \cite{1004}. However, for analytical simplicity and mathematical tractability, most works have considered single-antenna nodes and Rayleigh fading channels, and overlooked randomly distributed eavesdroppers. In order to investigate a more practical RIS-aided secure system, the randomness of the eavesdropper locations has to be taken into account when analyzing the system performance.

Stochastic geometry is an efficient mathematical tool for capturing the topological randomness of networks \cite{115}. In this approach, the wireless network is conveniently abstracted to a point process that can capture the essential network properties. A homogeneous Poisson point process (PPP) is the most popular and tractable point process used to model the locations of the mobile devices in wireless networks \cite{1005}. However, there have been few works studying the secrecy performance of an RIS-aided communication system with spatially random eavesdroppers, which leaves the impact of key system parameters under the stochastic geometry framework still unclear.

\subsection{Motivation and Contribution}
{PLS has been studied in diverse RIS-assisted communication scenarios, but rarely considered in the general case of randomly located eavesdroppers. Although the authors of \cite{4} and \cite{5} considered random eavesdropper locations, there are still several research gaps left to be filled. In \cite{4}, a single-antenna setting was adopted at the base station to analyze the ESC performance, but the Rician fading assumption and RIS phase shifts optimization were not taken into consideration for the multiple-antenna setting. In \cite{5}, a study was developed based on a simplified transmit beamforming design and the analyses of secrecy diversity order and capacity were not conducted. In addition, the works in \cite{4} and \cite{5} both need to be further explored to quantify the impact of the key parameters, e.g., the number of RIS reflecting elements, on the attainable secrecy performance in order to provide insightful guidelines for system design.} 
{Table \ref{tabel1} provides a summary of current studies related to the RIS-assisted secure communication systems and compares our work with them.} 
{In this paper, we investigate the SOP and ESC performance of an RIS-assisted multiple-input single-output (MISO) system over Rician fading channels with spatially random eavesdroppers. The main contributions of our work are summarized as follows.}

\renewcommand\arraystretch{1}
\begin{table*}[t]
  \caption{{Comparison between our work with the state-of-the-art.}}
  \begin{center}
  \setlength{\tabcolsep}{3.5mm}{
    \begin{tabular}{|c|c|c|c|c|c|c|c|c|c|c|}
      \hline
       & {\cite{111}} & {\cite{3}} & {\cite{112}} & {\cite{113}} & {\cite{114}} & {\cite{1003}} & {\cite{1004}} & {\cite{4}} & {\cite{5}} & {Our work}\\
      \hline
      MISO &  &  &  &  &  &  &  & {$\checkmark$} & {$\checkmark$} & {$\checkmark$} \\
      \hline
      {Rician channel} &  &  &  &  &  &  &  &  & {$\checkmark$} & {$\checkmark$} \\
      \hline
      {Eves with PPP distribution} &  &  &  &  &  &  &  & {$\checkmark$} & {$\checkmark$} & {$\checkmark$} \\
      \hline
      {SOP analysis} & {$\checkmark$} & {$\checkmark$} &  & {$\checkmark$} & {$\checkmark$}  &  {$\checkmark$}  &  &  & {$\checkmark$} & {$\checkmark$} \\
      \hline
      {ESC analysis} &  &  & {$\checkmark$} &  &  &  & {$\checkmark$} & {$\checkmark$} &  & {$\checkmark$} \\
      \hline
      {High-SNR analysis} &  & {$\checkmark$} &  & {$\checkmark$} &  & {$\checkmark$} &  &  &  & {$\checkmark$} \\
      \hline
    \end{tabular}}
  \end{center}
  \label{tabel1}
\end{table*} 

\begin{itemize}
\item{{Assuming maximum ratio transmission (MRT) for the transmit beamforming, the optimal reflect beamforming at the RIS is designed. Using tools from stochastic geometry, we derive accurate closed-form expressions for the distributions of the received signal-to-noise-ratios (SNRs) at the legitimate user and randomly located eavesdroppers located according to a homogeneous PPP.}}
\end{itemize} 

\begin{itemize}
\item{{We propose a new analytical PLS framework for the RIS-aided MISO secure communication system over Rician fading channels. Novel closed-form expressions are derived to characterize the SOP and ESC of the system. The derived results show that both the SOP and ESC are mainly affected by the number of RIS reflecting elements, and are not strong functions of the number of transmit antennas nor the transmit power at the base station, which means that increasing either of these latter resources will not significantly improve the secrecy performance.}}
\end{itemize} 

\begin{itemize}
\item{{To obtain more insightful observations, the asymptotic secrecy performance in the high SNR regime is also analyzed to characterize the SOP and ESC. The asymptotic SOP demonstrates that the secrecy diversity order of RIS-aided MISO secure communication systems depends on the path loss exponent of the RIS-to-ground links. In addition, when the locations of spatially random eavesdroppers are unknown, it is more efficient to deploy the RIS closer to the legitimate user than to the base station. Moreover, the impact of the randomly located eavesdropper density, $\lambda_e$, on the asymptotic ESC performance is quantitatively evaluated to be additive and proportional to $\log_2{\left({1}/{\lambda_e}\right)}$.}}
\end{itemize} 

\subsection{Organization}
The rest of this paper is organized as follows. We introduce the system model in Section~\uppercase\expandafter{\romannumeral2}, and in Section \uppercase\expandafter{\romannumeral3}, we derive the exact distributions of the received SNRs for the legitimate user and randomly located eavesdroppers. In Section \uppercase\expandafter{\romannumeral4}, we provide a closed-form expression for the SOP, and we study the secrecy diversity order at high SNR. The ergodic secrecy capacity is investigated in Section \uppercase\expandafter{\romannumeral5} and simulation results are discussed in Section \uppercase\expandafter{\romannumeral6}. Finally, we draw our conclusions in Section \uppercase\expandafter{\romannumeral7}.

\emph{Notation:} Boldface lowercase (uppercase) letters represent vectors (matrices). The set of all complex numbers is denoted by $\mathbb{C}$. The set of all positive real numbers is denoted by $\mathbb{Z}^+$. The superscripts $(\cdot)^T$, $(\cdot)^\ast$, and $(\cdot)^H$ stand for the transpose, conjugate, and conjugate-transpose operations, respectively. A circularly symmetric complex Gaussian distribution is denoted by ${\cal {CN}}$. A Rician distribution is denoted by $Rice$. $\mathbb{E}\{\cdot\}$ denotes the expectation of a random variable (RV). ${\rm diag}\left\{\cdot\right\}$ indicates a diagonal matrix. The operators $|\cdot|$ and $\left\|\cdot\right\|$ take the norm of a complex number and a vector, respectively. $\angle$ returns the phase of a complex value. The symbol $\Gamma\left(\cdot\right)$ is the Gamma function. The symbols $\gamma\left(\cdot,\cdot\right)$ and $\Gamma\left(\cdot,\cdot\right)$ denote the lower and upper incomplete Gamma functions, respectively. $\left[x\right]^+={\rm max}\left\{0,x\right\}$ returns the maximum between $0$ and $x$.


\section{System Model}
\begin{figure}[!t]
\centering
\includegraphics[width=3in]{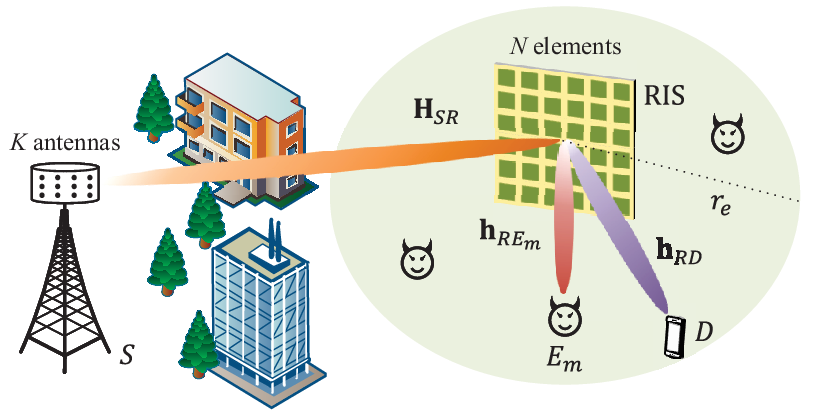}
\caption{System model of an RIS-assisted MISO communication system in the presence of spatially
random eavesdroppers.}
\label{fig1}
\end{figure}

As illustrated in Fig.~\ref{fig1}, an RIS-assisted secure communication system is considered, where a base station ($S$) is equipped with $K$ antennas and an RIS is composed of $N$ reflecting elements.
The reflection matrix of the RIS is denoted by $\mathbf{\Theta}\triangleq {\rm diag}\left\{{\eta_1}{\rm e}^{j\theta_1},\ldots,{\eta_n}{\rm e}^{j\theta_n},\ldots,{\eta_{N}}{\rm e}^{j\theta_N}\right\}$, where $\eta_n\in[0,1]$ and $\theta_n\in[0,2\pi)$ for $n=1, 2, \ldots, N$ are, respectively, the amplitude coefficient and the phase shift introduced by the $n$-th reflecting element. In order to exploit the maximum reflection capability of the RIS, the amplitude coefficients in this paper are set to 1, i.e., $\eta_{n}\!=\!1$ for all $n$.
{The eavesdroppers ($E$) are randomly located within a disk of radius $r_e$ centered at the RIS, and their spatial distribution is modeled using a homogeneous PPP $\Phi_e$ with a density $\lambda_{e}$ \cite{1005,4,5}. In contrast, the legitimate user is located without any spatial restrictions.}

{We assume that the direct link between $S$ and the legitimate user ($D$) is blocked by obstacles, which is a common occurrence in high frequency bands. To address this issue, the data transmission between $S$ and $D$ is facilitated by the RIS. Since the base station and the RIS are typically deployed at an elevated height with few scatterers, the channel between $S$ and the RIS is assumed to obey a line-of-sight (LoS) model \cite{116,116-2,117}, denoted by $\mathbf{H}_{S\!R}\in\mathbb{C}^{N\times K}$. While the legitimate user and eavesdroppers are usually located on the ground, the RIS-related channels with these terminals undergo both direct LoS and rich scattering, which can be modeled using Rician fading.} Here, $\mathbf{h}_{Ri}\in\mathbb{C}^{N\times1}$ is the channel vector of the RIS-$i$ links, where $i\in\left\{D,E_m\right\}$ and $E_m$ represents the $m$-th eavesdropper. Specifically, the expressions for $\mathbf{H}_{S\!R}$ and $\mathbf{h}_{Ri}$ are given by
\begin{align}
\mathbf{H}_{S\!R}=\sqrt\nu {\overline{\mathbf{H}}}_{S\!R},
\label{equal1}
\end{align}
\begin{align}
\mathbf{h}_{Ri}=\sqrt{\mu_i}\left(\sqrt{\frac{\epsilon}{\epsilon+1}}{\overline{\mathbf{h}}}_{Ri}+\sqrt{\frac{1}{\epsilon+1}}{\widetilde{\mathbf{h}}}_{Ri}\right),
\label{equal2}
\end{align}
where $\nu=\beta_0d_{S\!R}^{-\alpha_1}$ and $\mu_i=\beta_0d_{Ri}^{-\alpha_2}$ denote the large-scale fading coefficients, $\beta_0$ is the path loss at a reference distance of $1$m, $d_{S\!R}$ and $d_{Ri}$ are the distances of the $S$-RIS and RIS-$i$ links respectively, {$\alpha_1\geq2$ and $\alpha_2\geq2$ are the path loss exponents for the $S$-RIS and RIS-$i$ links respectively, and $\epsilon$ denotes the Rician factor. \footnote{{The Rician factors/path loss exponents for the RIS-$D$ and RIS-$E$ channels are assumed to be identical, as shown in \cite{5}\cite{1001}\cite{1002}, since the propagation environments around the legitimate user and eavesdroppers are similar.}}} The vector ${\widetilde{\mathbf{h}}}_{Ri}$ represents the non-line-of-sight (NLoS) component, whose entries are standard independent and identically distributed (i.i.d.) Gaussian RVs, i.e., $\mathcal{CN}(0,1)$. The LoS components ${\overline{\mathbf{H}}}_{S\!R}$ and ${\overline{\mathbf{h}}}_{Ri}$ are expressed as
\begin{align}
{\overline{\mathbf{H}}}_{S\!R}=\mathbf{a}_N\left(\phi_{S\!R}^a,\phi_{S\!R}^e\right)\mathbf{a}_K^H\left(\psi_{S\!R}^a,\psi_{S\!R}^e\right)=\mathbf{a}_{N,S\!R}\mathbf{a}_{K,S\!R}^H,
\label{equal3}
\end{align}
\begin{align}
{\overline{\mathbf{h}}}_{Ri}=\mathbf{a}_N\left(\psi_{Ri}^a,\psi_{Ri}^e\right)=\mathbf{a}_{N,Ri},
\label{equal4}
\end{align}
where $\phi_{S\!R}^a$ ($\phi_{S\!R}^e$) is the azimuth (elevation) angle of arrival (AoA) at the RIS, $\psi_{S\!R}^a$ ($\psi_{S\!R}^e$) and $\psi_{Ri}^a$ ($\psi_{Ri}^e$) are the azimuth (elevation) angles of departure (AoD) at the base station and RIS, respectively, and $\mathbf{a}_Z\left(\vartheta^a,\vartheta^e\right)$ is the array response vector. We consider a uniform square planar array (USPA) deployed at both the base station and the RIS. Thus, the array response vector can be written as \cite{6}
\begin{align}
\mathbf{a}_Z\left(\vartheta^a,\vartheta^e\right)&\!=\!\left[1,\ldots,{\rm e}^{j2\pi\frac{d}{\lambda}\left(x\sin{\vartheta^a}\sin{\vartheta^e}+y\cos{\vartheta^e}\right)},\ldots,\right.\nonumber\\
&\left.{\rm e}^{j2\pi\frac{d}{\lambda}\left(\left(\sqrt Z\!-\!1\right)\sin{\vartheta^a}\sin{\vartheta^e}+\left(\sqrt Z\!-\!1\right)\cos{\vartheta^e}\right)}\right]^T,
\label{equal5}
\end{align}
where $d$ and $\lambda$ are the element spacing and signal wavelength, respectively, and $0\le x,y<\sqrt Z$ are the element indices in the plane. We assume that the channel coefficients of $\mathbf{H}_{S\!R}$ and $\mathbf{h}_{R\!D}$ are perfectly known to $S$ and extensive approaches have been proposed in literature for the channel estimation of RIS-aided links \cite{300,301,302}. However, the channel coefficient $\mathbf{h}_{R\!E_m}$ is typically not available to $S$, since eavesdroppers are usually passive devices that do not emit signals.

\section{Distributions of the Received SNRs}
In order to analyze the secrecy performance of the system, we need to first characterize the distributions of the received SNRs at the legitimate user and the eavesdroppers.
\subsection{Distribution of the Received SNR at $D$}
Assuming quasi-static flat fading channels, the signal received at $D$ is expressed as
\begin{equation}
r_D=\mathbf{g}_D^H\mathbf{f}s+n_D,
\label{equal6}
\end{equation}
where the cascaded channel $\mathbf{g}_D^H\triangleq\mathbf{h}_{R\!D}^H\mathbf{\Theta}\mathbf{H}_{S\!R}\in\mathbb{C}^{1\times K}$, $\mathbf{f}\in\mathbb{C}^{K\times1}$ is the normalized beamforming vector, $s$ denotes the transmit signal that satisfies the power constraint $\mathbb{E}\{\left|s\right|^2\}=P_T$, and $n_D\sim\mathcal{CN}(0,\sigma_D^2)$ is additive white Gaussian noise (AWGN) at $D$ with variance $\sigma_D^2$. 
Therefore, the received SNR at $D$ is calculated as
\begin{equation}
\gamma_D=\frac{P_T\left|\mathbf{g}_D^H\mathbf{f}\right|^2}{\sigma_D^2}=\rho_d\left|{\rm A}\right|^2,
\label{equal7}
\end{equation}
where the RV $\left|{\rm A}\right|\triangleq\left|\mathbf{g}_D^H\mathbf{f}\right|$, and $\rho_d\triangleq\frac{P_T}{\sigma_D^2}$ denotes the transmit SNR from $S$ to $D$.

{Due to passive eavesdropping, the channel state information of the eavesdropper is not known to the RIS, thus we determine the transmit beamforming at $S$ and the reflect beamforming at the RIS by maximizing the received signal power at $D$.} 

\emph{Theorem~1:} When MRT beamforming is adopted, i.e., $\mathbf{f}=\frac{\mathbf{g}_D}{\left\|\mathbf{g}_D^H\right\|}$, the optimal reflection matrix of the RIS is given as
\begin{equation}
\mathbf{\Theta}^{\rm opt}={\rm diag}\left\{{\rm e}^{-j\angle\left({\rm diag}\left\{\mathbf{h}_{R\!D}^H\right\}\mathbf{a}_{N\!,S\!R}\right)}\right\}.
\label{equal8}
\end{equation}

\emph{Proof:} See Appendix A.$\hfill\blacksquare$

With the optimized RIS phase shifts in \emph{Theorem~1}, the RV $\left|{\rm A}\right|$ is expressed as $\left|{\rm A}\right|\!=\!\sqrt{K\nu}\sum_{n=1}^{N}\left|h_{R\!D}\left(n\right)\right|$ which follows the distribution characterized in the following lemma.

\emph{Lemma~1:} The cumulative distribution function (CDF) of $\left|{\rm A}\right|$ is well approximated by
\begin{equation}
F_{\left|{\rm A}\right|}\left(x\right)=\frac{1}{\Gamma\left(k\right)}\gamma\left(k,\frac{x}{\theta}\right),
\label{equal9}
\end{equation}
with shape parameter $k=N\frac{\frac{\pi}{4}\left(L_\frac{1}{2}\left(-\epsilon\right)\right)^2}{1+\epsilon-\frac{\pi}{4}\left(L_\frac{1}{2}\left(-\epsilon\right)\right)^2}$ and scale parameter $\theta=\sqrt K\sqrt{\frac{\mu_D\nu}{\epsilon+1}}\frac{1+\epsilon-\frac{\pi}{4}\left(L_\frac{1}{2}\left(-\epsilon\right)\right)^2}{\frac{\sqrt\pi}{2}L_\frac{1}{2}\left(-\epsilon\right)}$, in which $L_q\left(x\right)$ is the Laguerre polynomial defined in \cite[Eq.~(2.66)]{7}.

\emph{Proof:} See Appendix B.$\hfill\blacksquare$

By applying \emph{Lemma~1}, we can obtain the CDF and probability density function (PDF) of $\gamma_{D}$ in (\ref{equal7}), respectively, as
\begin{equation}
F_{\gamma_D}\left(x\right)=F_{\left|{\rm A}\right|}\left(\sqrt{{x}/{\rho_d}}\right)=\frac{1}{\Gamma\left(k\right)}\gamma\left(k,\frac{\sqrt{{x}/{\rho_d}}}{\theta}\right),
\label{equal10}
\end{equation}
\begin{equation}
f_{\gamma_D}\left(x\right)=\frac{dF_{\gamma_D}\left(x\right)}{dx}=\frac{{\rm e}^{-\frac{\sqrt{{x}/{\rho_d}}}{\theta}}\left(\frac{\sqrt{{x}/{\rho_d}}}{\theta}\right)^k}{2\Gamma\left(k\right)x}.
\label{equal11}
\end{equation}

\emph{Corollary~1 (Asymptotic Analysis):} For large $N$, the average received SNR at the legitimate user $D$ is obtained as
\begin{equation}
\mathbb{E}\{\gamma_D\}=\frac{\pi\left(L_{1/2}\left(-\epsilon\right)\right)^2}{4\left(\epsilon+1\right)}{\rho_d\mu_D\nu}KN^2.
\label{equal11-1}
\end{equation}

\emph{Proof:} We can infer from (\ref{equal7}) and \emph{Lemma~1} that $\mathbb{E}\{\gamma_D\}={\rho_d}\mathbb{E}\{|{\rm A}|^2\}={\rho_d}k(1+k){\theta}^2$ \cite{18}. By substituting the shape and
scale parameters, the proof is completed. $\hfill\blacksquare$

{\emph{Remark~1:} From \emph{Corollary~1}, we see that $\mathbb{E}\{\gamma_D\}$ scales with $K$ and $N^2$, which implies that deploying more transmit antennas and RIS reflecting elements both increase the average received SNR at the legitimate user, while the impact of $N$ is more dominant. Such a ``squared improvement" in terms of $N$ is due to the fact that the optimal reflect beamforming attained in \emph{Theorem~1} not only enables the system to achieve a beamforming gain of $KN$ in the $S$-RIS link, but also acquires an additional gain of $N$ by coherently collecting signals in the RIS-$D$ link.} 

{\emph{Remark~2:} From \emph{Corollary~1}, it is also found that $\mathbb{E}\{\gamma_D\}$ is increasing with respect to (w.r.t.) the Rician factor $\epsilon>0$. In other words, the average received SNR at the legitimate user is greater when the proportion of the LoS component in the RIS-$D$ link is higher. It is worth noting that the average SNR in (\ref{equal11-1}) can be upper bounded as $\mathbb{E}\{\gamma_D\}\le\mathbb{E}\{\gamma_D\}^{\rm U}={\rho_d}\mu_D\nu KN^2$ which holds for large $N$ and $\epsilon$ since $\lim_{\epsilon\to\infty}{\frac{\left(L_{1/2}\left(-\epsilon\right)\right)^2}{\epsilon+1}}=\frac{4}{\pi}$.}

\subsection{Distribution of the Received SNR at $E$}
Before calculating the effective SNR of the independent and homogeneous PPP distributed eavesdroppers, we first derive the SNR of the $m$-th eavesdropper $E_m$. The signal received at $E_m$ is formulated as
\begin{equation}
r_{E_m}=\mathbf{h}_{R\!E_m}^H\mathbf{\Theta}\mathbf{H}_{S\!R}\mathbf{f}s+n_{E_m},
\label{equal12}
\end{equation}
where $n_{E_m}\sim\mathcal{CN}(0,\sigma_E^2)$ is AWGN at $E_m$ with variance $\sigma_E^2$.
The received SNR at $E_m$ is given in the following proposition.

\emph{Proposition~1:} The received SNR at $E_m$ is expressed as
\begin{equation}
\gamma_{E_m}=\rho_eK\nu\left|Z_{E_m}\right|^2,
\label{equal13}
\end{equation}
where $\rho_e\triangleq\frac{P_T}{\sigma_E^2}$ denotes the transmit SNR and we define the RV $Z_{E_m}\!\triangleq\!\sum_{n=1}^{N}{h_{R\!E_m}^\ast\left(n\right){\rm e}^{-j\angle h_{R\!D}^\ast\left(n\right)}}$.

\emph{Proof:} See Appendix C.$\hfill\blacksquare$

According to \emph{Proposition~1}, we present \emph{Lemma~2} before deriving the distribution of $\gamma_{E_m}$.

\emph{Lemma~2:} The RV $Z_{E_m}$ follows a complex Gaussian distribution with mean $M_{E_m}$ and variance $V_{E_m}$, given by
\begin{align} 
~M_{E_m}=&\sqrt{\frac{\mu_{E_m}\epsilon^2}{{\frac{\pi}{4}\left(\epsilon+1\right)\left(L_\frac{1}{2}\left(-\epsilon\right)\right)}^2}}{\rm e}^{j\pi\frac{d}{\lambda}\left(\sqrt N-1\right)\left(\delta_1+\delta_2\right)}\nonumber\\
&\times\frac{\sin{\left(\pi\frac{d}{\lambda}\sqrt N \delta_1\right)}\sin{\left(\pi\frac{d}{\lambda}\sqrt N \delta_2\right)}}{\sin{\left(\pi\frac{d}{\lambda}\delta_1\right)}\sin{\left(\pi\frac{d}{\lambda}\delta_2\right)}},
\end{align}
\begin{equation} 
V_{E_m}=N\mu_{E_m}\left[1-\frac{\epsilon^2}{\frac{\pi}{4}\left(\epsilon+1\right){\left(L_\frac{1}{2}\left(-\epsilon\right)\right)}^2}\right], 
\end{equation}
where $\delta_1\!=\!\sin{\psi_{R\!D}^a}\sin{\psi_{R\!D}^e}\!-\!\sin{\psi_{R\!E_m}^a}\sin{\psi_{R\!E_m}^e}$ and $\delta_2\!=\!\cos{\psi_{R\!D}^e}\!-\!\cos{\psi_{R\!E_m}^e}$.

\emph{Proof:} See Appendix D.$\hfill\blacksquare$

As disclosed in \emph{Lemma~2}, we conclude that $\gamma_{E_m}$ is a non-central Chi-squared RV with two degrees of freedom. Then, the CDF of $\gamma_{E_m}$ is given by
\begin{equation}
F_{\gamma_{E_m}}\left(x\right)=1-Q_1\left(\frac{s}{\sigma},\frac{\sqrt x}{\sigma}\right),
\label{equal14}
\end{equation}
where $Q_1(\cdot,\cdot)$ is the first-order Marcum $Q$-function \cite{9}, and
\begin{align} 
\!\!\!\!s\!=\!\!\sqrt{\frac{\rho_eK\nu\mu_{E_m}\epsilon^2}{\!\frac{\pi}{4}\!\left(\!\epsilon\!+\!1\!\right){\!\left(\!L_\frac{1}{2}\!\left(\!-\epsilon\right)\right)}^2}}\!\left|\!\frac{\sin\!{\left(\!\pi\frac{d}{\lambda}\sqrt N \delta_1\!\right)}\!\sin{\!\left(\!\pi\frac{d}{\lambda}\sqrt N \delta_2\!\right)}}{\sin\!{\left(\pi\frac{d}{\lambda}\delta_1\right)}\!\sin\!{\left(\pi\frac{d}{\lambda}\delta_2\right)}}\!\right|,
\end{align}
\begin{align}
\sigma^2\!=\!\frac{1}{2}\rho_eKN\nu\mu_{E_m}\left[1-\frac{\epsilon^2}{\frac{\pi}{4}\left(\epsilon+1\right){\left(L_\frac{1}{2}\left(-\epsilon\right)\right)}^2}\right].
\end{align} 

In the case of non-colluding eavesdroppers, the eavesdropper with the strongest channel dominates the secrecy performance. Thus, the corresponding CDF of the eavesdropper SNR is derived as
\begin{align}
F_{\gamma_E}\left(x\right)&={\rm Pr}\left\{\mathop{\max}\limits_{m\in{\Phi_e}}{\gamma_{E_m}\le x}\right\}\nonumber\\
&\mathop=^{\left(\rm a\right)}\mathbb{E}_{\Phi_e}\left\{\prod_{m\in\Phi_e,r_m\le r_e}{F_{\gamma_{E_m}}\left(x\right)}\right\}\nonumber\\
&\mathop=^{\left(\rm b\right)}{\rm exp}\left[-2\pi\lambda_e\int_{0}^{r_e}{\left(1-F_{\gamma_{E_m}}\left(x\right)\right)r\,{\rm{d}}r}\right]\nonumber\\
&\mathop=^{\left(\rm c\right)}{\rm exp}\left[-2\pi\lambda_e\int_{0}^{r_e}{Q_1\left(\varpi,\Xi\sqrt x r^\frac{\alpha_2}{2}\right)r\,{\rm{d}}r}\right],
\label{equal15}
\end{align}
where $({\rm a})$ follows from the i.i.d. characteristic of the eavesdroppers' SNRs and their independence from the point process ${\Phi_e}$, $({\rm b})$ follows from the probability generating functional (PGFL) of the PPP \cite[Eq.~(4.55)]{8}, and $({\rm c})$ is obtained by using $\mu_{E_m}=\beta_0r^{-\alpha_2}$ and defining the parameters
\begin{align}
\varpi\triangleq&\sqrt2\left|\frac{\sin{\left(\pi\frac{d}{\lambda}\sqrt N \delta_1\right)}\sin{\left(\pi\frac{d}{\lambda}\sqrt N \delta_2\right)}}{\sin{\left(\pi\frac{d}{\lambda}\delta_1\right)}\sin{\left(\pi\frac{d}{\lambda}\delta_2\right)}}\right|\nonumber\\
&\times\left[N\left(\frac{\frac{\pi}{4}\left(\epsilon+1\right){\left(L_\frac{1}{2}\left(-\epsilon\right)\right)}^2}{\epsilon^2}-1\right)\right]^{-\frac{1}{2}},
\end{align}
\begin{align}
\Xi\!\triangleq\!{\sqrt2}\!\left[NK\nu\beta_0\rho_e\!\left(\!1\!-\!\frac{\epsilon^2}{\frac{\pi}{4}\!\left(\epsilon\!+\!1\right)\!{\left(L_\frac{1}{2}\left(-\epsilon\right)\right)}^2}\right)\right]^{-\frac{1}{2}}.
\end{align}

From the characterization in \cite[Eq.~(2)]{9}, we have the following approximation for the Marcum $Q$-function in (\ref{equal15}): $Q_1\left(\varpi,\Xi\sqrt x r^\frac{\alpha_2}{2}\right)\simeq{\rm exp}\left[-{\rm e}^{v\left(\varpi\right)}\left(\Xi\sqrt x r^\frac{\alpha_2}{2}\right)^{\mu\left(\varpi\right)}\right]$, where $v\left(\varpi\right)$ and $\mu\left(\varpi\right)$ are polynomial functions of $\varpi$ defined as $v\left(\varpi\right)\!=\!-0.840\!+\!0.327\varpi\!-\!0.740\varpi^2\!+\!0.083\varpi^3\!-\!0.004\varpi^4$ and $\mu\left(\varpi\right)\!=\!2.174\!-\!0.592\varpi\!+\!0.593\varpi^2\!-\!0.092\varpi^3\!+\!0.005\varpi^4$. Then, (\ref{equal15}) is further calculated as
\begin{align}
F_{\gamma_E}\left(x\right)&\!=\!{\rm exp}\left[\!-\!2\pi\lambda_e\!\int_{0}^{r_e}\!{{\rm exp}\!\left[-{\rm e}^{v\left(\varpi\right)}\left(\Xi\sqrt x r^\frac{\alpha_2}{2}\right)^{\mu\left(\varpi\right)}\right]\!r{\rm{d}}r}\right]\nonumber\\
&\!=\!{\rm exp}\left[\!-t_0\frac{\Gamma\left(t_1\right)-\Gamma\left(t_1,t_2x^{t_3}\right)}{x^{t_4}}\right],
\label{equal16}
\end{align}
where the last equality is obtained from \cite[Eq.~(3.326)]{10} with the definitions $t_0=\frac{2\pi\lambda_e}{\frac{\alpha_2}{2}\mu\left(\varpi\right){\rm e}^\frac{4v\left(\varpi\right)}{\alpha_2\mu\left(\varpi\right)}\Xi^\frac{4}{\alpha_2}}$, $t_1=\frac{2}{\frac{\alpha_2}{2}\mu\left(\varpi\right)}$, $t_2={\rm e}^{v\left(\varpi\right)}\Xi^{\mu\left(\varpi\right)}{r_e}^{\frac{\alpha_2}{2}\mu\left(\varpi\right)}$, $t_3=\frac{\mu\left(\varpi\right)}{2}$, and $t_4=\frac{2}{\alpha_2}$.

Therefore, the PDF of the overall eavesdropper SNR can be further derived from (\ref{equal16}) as
\begin{align}
f_{\gamma_E}\left(x\right)=&\frac{dF_{\gamma_E}\left(x\right)}{dx}=t_0x^{-t_4-1}\left(t_4\gamma\left(t_1,t_2x^{t_3}\right)\right.\nonumber\\
&\left.-t_3\left(t_2x^{t_3}\right)^{t_1}{\rm e}^{-t_2x^{t_3}}\right){\rm e}^{-t_0x^{-t_4}\gamma\left(t_1,t_2x^{t_3}\right)}.
\label{equal17}
\end{align}





\section{Secrecy Outage Analysis}
In this section, we apply the derived statistical properties of $\gamma_D$ and $\gamma_E$ in the above section to conduct the secrecy outage analysis of the RIS-aided MISO system. 
\subsection{Theoretical SOP Analysis}
A popular metric for quantifying PLS is the SOP, which is defined as the probability that the instantaneous secrecy capacity falls below a target secrecy rate $C_{\rm th}$. Mathematically, the SOP is evaluated by
\begin{align}
{\rm SOP}&={\rm Pr}\left(\ln{\left(1+\gamma_D\right)}-\ln{\left(1+\gamma_E\right)}<C_{\rm th}\right)\nonumber\\
&=\int_{0}^{\infty}{F_{\gamma_D}\left(\left(1+x\right)\varphi-1\right)}f_{\gamma_E}\left(x\right){\rm{d}}x,
\label{equal18}
\end{align}
where $\varphi\triangleq {\rm e}^{C_{\rm th}}$. By substituting (\ref{equal10}) and (\ref{equal17}) into (\ref{equal18}), the SOP can be easily expressed as follows
\begin{align}
{\rm SOP}=\frac{1}{\Gamma\left(k\right)}t_0\left(t_4I_1-t_3{t_2}^{t_1}I_2\right),
\label{equal19}
\end{align}
where $I_1$ and $I_2$ are defined as
\begin{align}
~~~~~I_1 =\int_{0}^{+\infty}&\gamma\left(k,\frac{\sqrt{\frac{\left(\left(1+x\right)\varphi-1\right)}{\rho_d}}}{\theta}\right)x^{-t_4-1}\nonumber\\
\times~&\gamma\left(t_1,t_2x^{t_3}\right){\rm e}^{-t_0x^{-t_4}\gamma\left(t_1,t_2x^{t_3}\right)}{\rm{d}}x,
\label{equal20}
\end{align}
\begin{align}
I_2 =\int_{0}^{+\infty}&\gamma\left(k,\frac{\sqrt{\frac{\left(\left(1+x\right)\varphi-1\right)}{\rho_d}}}{\theta}\right)x^{-t_4-1}x^{t_1t_3}\nonumber\\
\times~&{\rm e}^{-t_2x^{t_3}}{\rm e}^{-t_0x^{-t_4}\gamma\left(t_1,t_2x^{t_3}\right)}{\rm{d}}x.
\label{equal21}
\end{align}


{However, it is difficult to directly compute an accurate closed-form expression for (\ref{equal19}), because (\ref{equal20}) and (\ref{equal21}) both involve an intractable integral. In order to analyze the secrecy performance, an approximate closed-form expression for the SOP is presented in \emph{Proposition~2}.}

{\emph{Proposition~2:} When $r_e\rightarrow\infty$, the SOP can be approximated as\footnote{{The assumption of a large $r_e$, similar to \cite{20}\cite{13}, only denotes an upper limit of distance and does not mean that the eavesdroppers must be far away from the RIS. Also, we will illustrate that the insights we obtained under this assumption are still applicable when $r_e$ is relatively small in the simulations.}}}
\begin{align}
{\rm SOP}\simeq~&1-\frac{1}{\Gamma\left(k\right)}\frac{p^\frac{1}{2}q^{k-\frac{1}{2}}}{2^{\frac{p+4q}{2}-2k}\pi^{\frac{p+4q}{2}-1}}\nonumber\\
&\times G_{0,p+4q}^{p+4q,0}\left(\frac{\left(t_0\Gamma\left(t_1\right)\varphi^{t_4}\right)^p}{p^p\left(4q\sqrt{\rho_d}\theta\right)^{4q}}\middle|\begin{matrix}-\\\Delta\\\end{matrix}\right),
\label{equal22}
\end{align}
where $G_{s,t}^{m,n}\left(z\right)$ is Meijer's $G$ function \cite{10}, $p,q\in\mathbb{Z}^+$, ${p}/{q}=\alpha_2$, and $\Delta=\left[0,\frac{1}{p},\ldots,\frac{p-1}{p},\frac{k}{4q},\frac{k+1}{4q},\ldots,\frac{k+4q-1}{4q}\right]$.


\emph{Proof:} See Appendix E.$\hfill\blacksquare$

\emph{Proposition~2} provides an explicit relation between the secrecy outage probability and various system parameters. A number of interesting points can be made based on this expression.

{\emph{Remark~3:} The SOP in (\ref{equal22}) is not a function of the transmit power $P_T$ at the base station, which implies that increasing $P_T$ does not enhance the system's secrecy performance. This is intuitive since an increase in $P_T$ would yield a proportional increase in the transmit SNRs at both the legitimate user and eavesdroppers.}

\emph{Proof:} According to \emph{Proposition~2}, we see that the transmit power $P_T$ affects only the term $\frac{t_0^p}{\rho_d^{2q}}$ in (\ref{equal22}), which is given by
\begin{align}
\frac{t_0^p}{\rho_d^{2q}}&=\left(\frac{2\pi\lambda_e}{\frac{\alpha_2}{2}\mu\left(\varpi\right)e^\frac{4v\left(\varpi\right)}{\alpha_2\mu\left(\varpi\right)}}\right)^p\frac{1}{\left(\Xi^2\rho_d\right)^{2q}}\propto\left(\frac{\rho_e}{\rho_d}\right)^{2q}.
\label{equal22-1}
\end{align}
This equality shows that the SOP depends on the ratio of the transmit SNRs at the legitimate user and the eavesdroppers, i.e., $\frac{\rho_e}{\rho_d}$, regardless of the specific values of $P_T$. $\hfill\blacksquare$

{\emph{Remark~4:} The SOP in (\ref{equal22}) is mainly affected by the number of RIS reflecting elements, $N$, while the impact of the number of transmit antennas, $K$, is marginal. This is readily checked by calculating $\frac{t_0^p}{\theta^{4q}}$ in (\ref{equal22}) because the other terms are obviously irrelevant to $K$. We obtain $\frac{t_0^p}{\theta^{4q}}\!\!=\!\!\left(\frac{2\pi\lambda_e}{\frac{\alpha_2}{2}\mu\left(\varpi\right){\rm e}^\frac{4v\left(\varpi\right)}{\alpha_2\mu\left(\varpi\right)}}\right)^p\!\!\!\!\frac{1}{\left(\Xi\theta\right)^{4q}}$, where $\Xi\theta\!=\!\frac{\chi}{\sqrt N}$ depends only on $N$, since the coefficient $\chi$ is independent of $N$ and $K$.}

In addition, \emph{Proposition~2} is a general analysis of the SOP for any path loss exponent $\alpha_2$. Some specific case studies are reported below. Note that other values of $\alpha_2$ also admit closed-form expressions using (\ref{equal22}).

\emph{Corollary~2:} For the special case of $\alpha_2=2$, i.e., $p=2$ and $q=1$, which corresponds to free space propagation \cite{124}, the SOP in (\ref{equal22}) reduces to
\begin{align}
{\rm SOP}\simeq1-\frac{2^{k-1}}{\sqrt\pi\Gamma\left(k\right)}\ G_{0,3}^{3,0}\left(\frac{t_0\Gamma\left(t_1\right)\varphi}{4\rho_d\theta^2}\middle|\begin{matrix}-\\0,\frac{k}{2},\frac{k+1}{2}\\\end{matrix}\right).
\label{equal23}
\end{align}

\emph{Corollary~3:} For the special case of $\alpha_2=4$, i.e., $p=4$ and $q=1$, which is a common practical value for the path-loss exponent in outdoor urban environments \cite{124}, the SOP in (\ref{equal22}) simplifies to the following expression
\begin{align}
{\rm SOP}\simeq1\!-\!\frac{2}{\Gamma\left(k\right)}\!\left(\frac{t_0\Gamma\left(t_1\right)\sqrt\varphi}{\sqrt{\rho_d}\theta}\right)^\frac{k}{2}\!\!K_k\!\!\left(\!2\!\left(\frac{t_0\Gamma\left(t_1\right)\sqrt\varphi}{\sqrt{\rho_d}\theta}\right)^\frac{1}{2}\!\right),
\label{equal23-1}
\end{align}
where $K_\nu\left(\cdot\right)$ denotes the $\nu$-th-order modified Bessel function of the second kind \cite[Eq.~(8.407)]{10}.

From (\ref{equal23-1}), it is obvious that the SOP is a monotonically increasing function w.r.t. $\frac{t_0}{\sqrt{\rho_d}\theta}=\sqrt{\frac{\rho_e}{\rho_d}}\lambda_ed_{R\!D}^2\beta\left(N,\epsilon\right)$ with fixed $k$, where $\beta\left(N,\epsilon\right)$ consists of the parameters $N$, $\epsilon$, and constant terms. Some new observations can thus be obtained in addition to the results presented in \emph{Remark~3} and \emph{Remark~4}. We evince that the SOP increases with the density parameter $\lambda_e$, which implies that a larger density of randomly located eavesdroppers leads to a negative effect on the secrecy performance. Moreover, we can also see that the SOP is only related to the distance of the RIS-$D$ link, i.e., $d_{R\!D}$. Therefore, when the locations of the eavesdroppers are unknown, this suggests that the RIS should be deployed closer to the legitimate user than to the base station.

\subsection{Secrecy Diversity Order Analysis}
In order to derive the secrecy diversity order and gain further insights, we adopt the analytical framework proposed in \cite{11} where the secrecy diversity order is defined as follows
\begin{align}
d_s=-\lim_{\rho_d\rightarrow\infty}{\frac{\log{{\rm SOP}^\infty}}{\log{\rho_d}}},
\label{equal24}
\end{align}
where ${\rm SOP}^\infty$ represents the asymptotic value of the SOP in (\ref{equal22}) for $\rho_d\rightarrow\infty$, and the transmit SNR $\rho_e$ is fixed.

According to \cite[Eq.~(07.34.06.0006.01)]{16}, the SOP in (\ref{equal22}) can be expanded as
\begin{align}
{\rm SOP}\simeq&1-\frac{1}{\Gamma\left(k\right)}\frac{p^\frac{1}{2}q^{k-\frac{1}{2}}}{2^{\frac{p+4q}{2}-2k}\pi^{\frac{p+4q}{2}-1}}\times G_{0,p+4q}^{p+4q,0}\left(x\middle|\begin{matrix}-\\\Delta\\\end{matrix}\right) \nonumber\\
=&1-\frac{1}{\Gamma\left(k\right)}\frac{p^\frac{1}{2}q^{k-\frac{1}{2}}}{2^{\frac{p+4q}{2}-2k}\pi^{\frac{p+4q}{2}-1}}\times\nonumber\\
&\sum_{l=1}^{p+4q}{\prod_{j=1,j\neq l}^{p+4q}\!\Gamma\left(\Delta\left(j\right)\!-\!\Delta\left(l\right)\right)x^{\Delta\left(l\right)}}\left(1\!+\!\mathcal{O}\left(x\right)\right),
\label{equal25}
\end{align}
where $x\!=\!\!\frac{\left(t_0\Gamma\left(t_1\right)\varphi^{t_4}\right)^p}{p^p\left(4q\sqrt{\rho_d}\theta\right)^{4q}}\!\rightarrow\!0$, and $\mathcal{O}$ denotes higher order terms. 

When the transmit SNR from $S$ to $D$ is sufficiently large, i.e., $\rho_d\rightarrow\infty$, only the dominant terms $l=0$ and $l=1$ in the summation of (\ref{equal25}) are retained, which yields the asymptotic SOP as expressed in (\ref{equal27}),
\begin{figure*}[t] 
\begin{align}
{\rm SOP}^\infty=&1-\frac{1}{\Gamma\left(k\right)}\frac{p^\frac{1}{2}q^{k-\frac{1}{2}}}{2^{\frac{p+4q}{2}-2k}\pi^{\frac{p+4q}{2}-1}}\left[\prod_{j=2}^{p+4q}\Gamma\left(\Delta\left(j\right)\right)x^0+\prod_{j=1,j\neq2}^{p+4q}\Gamma\left(\Delta\left(j\right)-\frac{1}{p}\right)x^\frac{1}{p}\right] \nonumber\\
=&1-\frac{1}{\Gamma\left(k\right)}\frac{p^\frac{1}{2}q^{k-\frac{1}{2}}}{2^{\frac{p+4q}{2}-2k}\pi^{\frac{p+4q}{2}-1}}\left[\prod_{j=0}^{p-1}\Gamma\left(\frac{1}{p}+\frac{j}{p}\right)\prod_{j=0}^{4q-1}\Gamma\left(\frac{k}{4q}+\frac{j}{4q}\right)-p\prod_{j=0}^{p-1}\Gamma\left(\frac{1}{p}+\frac{j}{p}\right)\right.\nonumber\\
&\left.\times\prod_{j=0}^{4q-1}\Gamma\left(\frac{k}{4q}-\frac{1}{p}+\frac{j}{4q}\right)x^\frac{1}{p}\right]
=\frac{t_0\mathrm{\Gamma}\left(t_1\right)\varphi^\frac{2}{\alpha_2}\Gamma\left(k-\frac{4}{\alpha_2}\right)}{\theta^\frac{4}{\alpha_2}\Gamma\left(k\right)}\left(\rho_d\right)^{-\frac{2}{\alpha_2}},
\label{equal27}
\end{align}
\hrulefill
\end{figure*}
where the last step is calculated by applying Gauss' multiplication formula \cite[Eq.~(6.1.20)]{19}.
 
{\emph{Remark~5:} By substituting (\ref{equal27}) into (\ref{equal24}), the secrecy diversity order is obtained as $\frac{2}{\alpha_2}$, which only depends on the path loss exponent of the RIS-to-ground links. This implies that the secrecy diversity order of this system improves when the RIS is deployed to provide better LoS links to the terminals.}

\section{Ergodic Secrecy Capacity Analysis} 
In this section, we obtain closed-form expressions for both the theoretical and asymptotic ESC. We also characterize the impact of various parameters, including $N$, $K$, $P_T$, and $\lambda_e$, on the ESC performance of the system.

\subsection{Theoretical ESC Analysis}
The ESC is an alternative fundamental metric that denotes the statistical average of the secrecy rate over fading channels, which is mathematically expressed as
\begin{equation}
C_s=\mathbb{E}\left\{\left[\log_2{\left(1+\gamma_D\right)}-\log_2{\left(1+\gamma_E\right)}\right]^+\right\}.
\label{equal28}
\end{equation}

Given the received SNRs at both the legitimate user and eavesdroppers in Section \uppercase\expandafter{\romannumeral3}, we first derive an approximate expression for the ESC in the following proposition.

\emph{Proposition~3:} The ESC for the RIS-assisted system is evaluated as
\begin{align}
{\bar{C}}_s=\left[R_D-R_E\right]^+,
\label{equal29}
\end{align}
where $R_D$ and $R_E$ are the ergodic rates of the legitimate user $D$ and the eavesdroppers $E$, respectively, and are expressed as
\begin{align}
R_D=\frac{1}{\ln{2}}\frac{1}{\Gamma\left(k\right)}\frac{2^{k-\frac{1}{2}}}{\sqrt{2\pi}}G_{2,4}^{4,1}\left(\frac{1}{4\rho_d\theta^2}\middle|\begin{matrix}0,1\\0,0,\frac{k}{2},\frac{k+1}{2}\\\end{matrix}\right),
\label{equal30}
\end{align}
\begin{align}
~~R_E=\frac{1}{\ln{2}}\int_{0}^{+\infty}\frac{1-{\rm exp}\left[-t_0\frac{\Gamma\left(t_1\right)-\Gamma\left(t_1,t_2x^{t_3}\right)}{x^{t_4}}\right]}{1+x}{\rm{d}}x.
\label{equal31}
\end{align}

\emph{Proof:} Using Jensen's inequality, an effective approximation of the ESC can be written as \cite{15}\cite{123}
\begin{align}
\!\!{\bar{C}}_s\!=\!\left[\mathbb{E}\!\left\{\log_2\!{\left(1\!+\!\gamma_D\right)}\!-\!\log_2\!{\left(1\!+\!\gamma_E\right)}\right\}\right]^+\!=\!\left[R_D\!-\!R_E\right]^+,
\label{equal32}
\end{align}
where $R_D$ and $R_E$ are respectively calculated as follows
\begin{align}
R_D&=\mathbb{E}\!\left\{\log_2{\left(1+\gamma_D\right)}\right\}=\frac{1}{\ln{2}}\int_{0}^{+\infty}\frac{{\bar{F}}_{\gamma_D}\left(x\right)}{1+x}dx~~~~~~~~\nonumber\\
&=\frac{1}{\ln{2}}\int_{0}^{+\infty}\frac{\Gamma\left(k,\frac{\sqrt{{x}/{\rho_d}}}{\theta}\right)}{\Gamma\left(k\right)\left(1+x\right)}{\rm{d}}x,
\label{equal33}
\end{align}
\begin{align}
R_E&=\mathbb{E}\left\{\log_2{\left(1+\gamma_E\right)}\right\}=\frac{1}{\ln{2}}\int_{0}^{+\infty}\frac{{\bar{F}}_{\gamma_E}\left(x\right)}{1+x}dx\nonumber\\
&=\frac{1}{\ln{2}}\int_{0}^{+\infty}\frac{1-{\rm exp}\left[-t_0\frac{\Gamma\left(t_1\right)-\Gamma\left(t_1,t_2x^{t_3}\right)}{x^{t_4}}\right]}{1+x}{\rm{d}}x.
\label{equal34}
\end{align}

Using \cite[Eq.~(07.34.03.0613.01)]{16} and \cite[Eq.~(9.31.5)]{10}, (\ref{equal33}) is equivalently given by
\begin{align}
R_D&=\frac{1}{\ln{2}}\frac{1}{\Gamma\left(k\right)}\int_{0}^{+\infty}{\frac{1}{1+x}G_{1,2}^{2,0}\left(\frac{\sqrt x}{\sqrt{\rho_d}\theta}\middle|\begin{matrix}1\\k,0\\\end{matrix}\right)}{\rm{d}}x\nonumber\\
&=\frac{1}{\ln{2}}\frac{1}{\Gamma\left(k\right)}\frac{2^{k-\frac{1}{2}}}{\sqrt{2\pi}}G_{3,5}^{5,1}\left(\frac{1}{4\rho_d\theta^2}\middle|\begin{matrix}0,\frac{1}{2},1\\0,0,\frac{1}{2},\frac{k}{2},\frac{k+1}{2}\\\end{matrix}\right),
\label{equal35}
\end{align}
where the integral of (\ref{equal35}) is evaluated through \cite[Eq.~(07.34.21.0086.01)]{16}.

By further applying the identity given in \cite[Eq.~(9.31.1)]{10}, a simplified expression for the ergodic rate of the legitimate user $D$ is obtained in (\ref{equal30}). The proof is thus complete.$\hfill\blacksquare$

As a general analysis, the ergodic rate $R_E$ given by (\ref{equal31}) involves an intractable integral that is difficult to compute. So in the sequel we focus on a few relevant cases for widely-used values of path loss exponents where significant simplification is possible and intuitive ergodic secrecy rate expressions can be obtained.

\emph{Corollary~4:} For $r_e\rightarrow\infty$ and $\alpha_2=2$, which corresponds to the free space model \cite{124}, the ESC in (\ref{equal29}) reduces to the closed-form expression as
\begin{align}
{\bar{C}}_s=\left[R_D-R_{E,1}\right]^+,
\label{equal37}
\end{align}
where $R_D$ is given in closed-form by (\ref{equal30}) and
\begin{align}
R_{E,1}~=~&\frac{1}{\ln{2}}\int_{0}^{+\infty}\frac{1-{\rm exp}\left[-t_0\Gamma\left(t_1\right)x^{-1}\right]}{1+x}{\rm{d}}x\nonumber\\
=~&\frac{1}{\ln{2}}\left\{\gamma+\ln{\left(t_0\Gamma\left(t_1\right)\right)}+{\rm exp}\left[t_0\Gamma\left(t_1\right)\right]\right.\nonumber\\
&\left.\times\left({\rm Shi}\left(t_0\Gamma\left(t_1\right)\right)-{\rm Chi}\left(t_0\Gamma\left(t_1\right)\right)\right)\right\},
\label{equal38}
\end{align}
where $\gamma$ denotes Euler's constant, and ${\rm Shi}\left(\cdot\right)$ and ${\rm Chi}\left(\cdot\right)$ are the hyperbolic sine and cosine integral functions, respectively \cite[Eq.~(8.221)]{10}.

\emph{Corollary~5:} For $r_e\rightarrow\infty$ and $\alpha_2=4$, which is a common value for the path-loss exponent in outdoor urban environments \cite{124}, the ESC in (\ref{equal29}) reduces to the closed-form expression
\begin{align}
{\bar{C}}_s=\left[R_D-R_{E,2}\right]^+,
\label{equal39}
\end{align}
where $R_D$ is the same as (\ref{equal30}), and
\begin{align}
R_{E,2}&=\frac{1}{\ln{2}}\int_{0}^{+\infty}\frac{1-{\rm exp}\left[-t_0\Gamma\left(t_1\right)x^{-\frac{1}{2}}\right]}{1+x}{\rm{d}}x\nonumber\\
&=\frac{1}{\ln{2}}\frac{1}{\sqrt\pi}\ G_{2,4}^{3,2}\left(\frac{t_0^2{\Gamma\left(t_1\right)}^2}{4}\middle|\begin{matrix}1,1\\\frac{1}{2},1,1,0\\\end{matrix}\right).
\label{equal40}
\end{align}



\subsection{Asymptotic ESC Analysis}
In order to obtain useful insights for system design, we analyze the asymptotic ESC at high SNR in this subsection. For the sake of tractability, we first derive new expressions for bounding the ergodic rate $R_D$ in the following lemma.

{\emph{Lemma~3:} The ergodic rate $R_D$ can be upper bounded by}
{\begin{align}
\!\!\!\!R_D^{\rm U}\!=\!\log_2\!{\!\left(\!1\!+\!\rho_dKN\nu\mu_D\!\!\left(\!1\!+\!\!\frac{\pi}{4}\!\left(N\!-\!1\right)\frac{\left(\!L_{\frac{1}{2}}\!\left(\!-\epsilon\right)\!\right)^2}{\epsilon\!+\!1}\!\right)\!\right)}.
\label{equal43}
\end{align}}

{\emph{Proof:} See Appendix F.$\hfill\blacksquare$}

By applying \emph{Lemma~3}, we are ready to give the following corollary quantitatively analyzing the asymptotic ESC in the high SNR region.

\emph{Corollary~6:} In the high SNR regime, the ESC in (\ref{equal37}) is further simplified as follows
\begin{align}
{\bar{C}}_s\rightarrow&\left\{\log_2{\left(\frac{\sigma_E^2}{\sigma_D^2}\right)}+\log_2{\left(\frac{d_{RD}^{-2}}{\pi\lambda_e}\right)}-\frac{\gamma}{\ln{2}}+C\right.\nonumber\\
&+\left.\log_2{\left(\frac{1\!+\!\left(N-1\right)\frac{\pi}{4}\frac{1}{\epsilon+1}\left(L_\frac{1}{2}\left(-\epsilon\right)\right)^2}{1-\frac{\epsilon^2}{{\frac{\pi}{4}\left(\epsilon+1\right)\left(L_\frac{1}{2}\left(-\epsilon\right)\right)}^2}}\right)}\right\}^+,
\label{equal46}
\end{align}
where $C=\log_2\frac{\mu\left(\varpi\right){\rm e}^\frac{2v\left(\varpi\right)}{\mu\left(\varpi\right)}}{\Gamma\left(\frac{2}{\mu\left(\varpi\right)}\right)}$ is a constant.

\emph{Proof:} See Appendix G.$\hfill\blacksquare$

From \emph{Corollary~6}, we have the following remarks on the impact of key system parameters, including $N$, $K$, $P_T$, and $\lambda_e$, on the secrecy performance.

{\emph{Remark~6:} By direct inspection of (\ref{equal46}), it is evident that the asymptotic ESC does not depend on the number of transmit antennas, $K$, but it increases with the number of RIS reflecting elements, $N$. This suggests that deploying more RIS reflecting elements rather than transmit antennas achieves better secrecy performance. Moreover, for large $N$, the asymptotic ESC scales logarithmically with $N$.}

{\emph{Remark~7:} \emph{Corollary~6} also indicates that the asymptotic ESC is not a function of $d_{S\!R}$ whereas it increases with decreasing $d_{R\!D}$. This result might seem a bit counterintuitive at first, but it actually makes sense and can be explained as follows. When the distance from the base station to the RIS decreases, the received SNRs at both the legitimate user and the eavesdroppers increase by the same amount, and they offset each other. Therefore, if the locations of the eavesdroppers are unknown, this suggests that the RIS should be deployed closer to the legitimate user than to the base station.}

{\emph{Remark~8:} In agreement with {Remark~3}, the asymptotic ESC in (\ref{equal46}) is only related to the ratio of the noise power at the legitimate user and the eavesdroppers $\frac{\sigma_D^2}{\sigma_E^2}$, and is independent of the transmit power $P_T$. In addition, in line with intuition, the asymptotic ESC decreases when the density of the homogeneous PPP $\lambda_e$ increases. We see that the effect of $\lambda_e$ on the ESC is additive, and is proportional to $\log_2{\left({1}/{\lambda_e}\right)}$.}

\section{Numerical Results}
\begin{figure}[!t]
    \setlength{\abovecaptionskip}{0pt}
    \setlength{\belowcaptionskip}{0pt}
    \centering
    \includegraphics[width=95 mm,height=55mm]{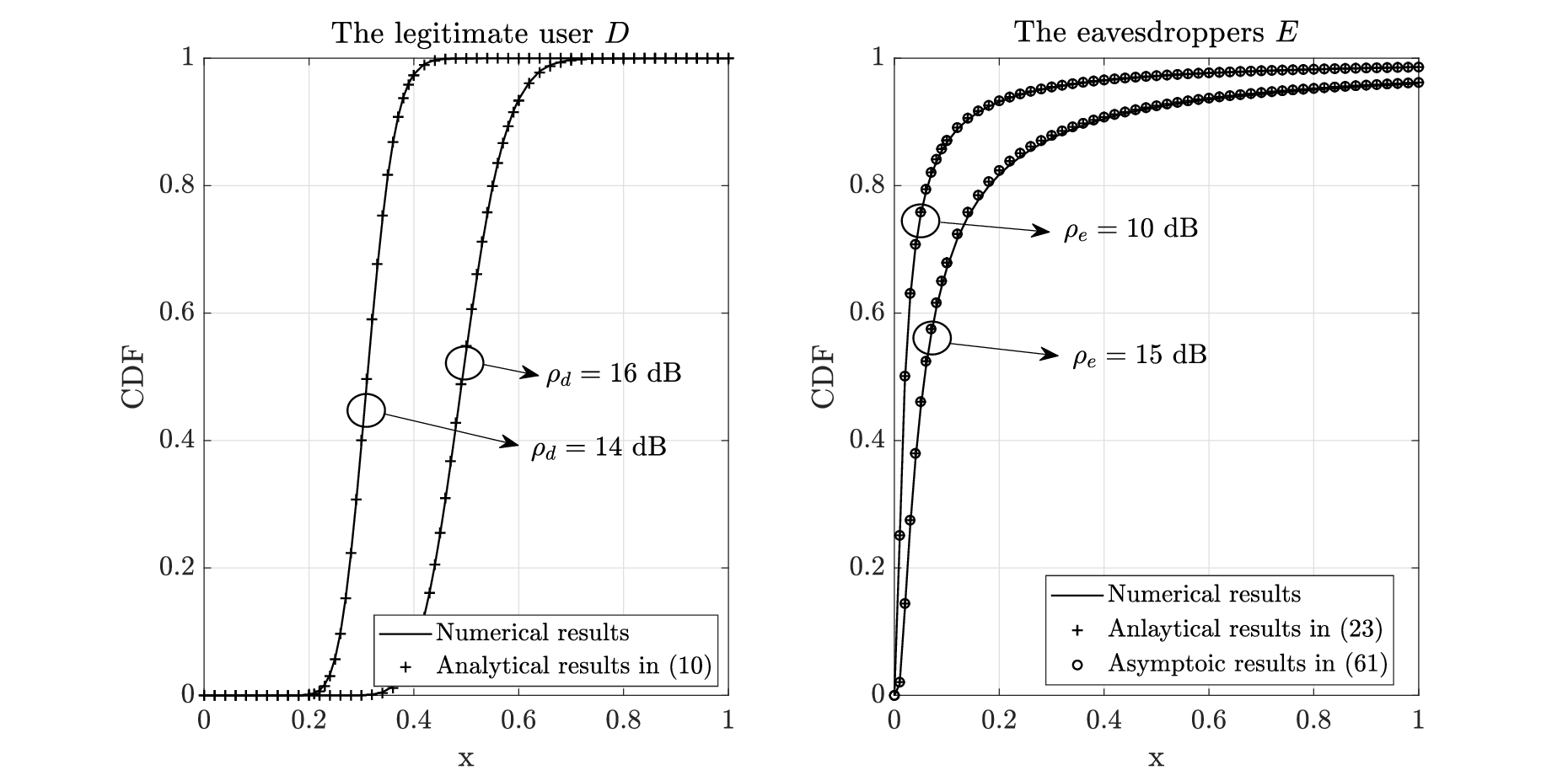}
    \caption{The CDF of the received SNRs at the legitimate user $D$ and the eavesdroppers $E$, respectively.}
    \label{fig2} \end{figure}

In this section, Monte-Carlo simulations are presented to validate the analytical results. All the simulation results are obtained by averaging over $10^5$ independent channel realizations.
We first verify the approximations of the received SNR distributions at the legitimate user and the eavesdroppers in Fig.~\ref{fig2}. The simulation parameters are set to $K=16$, $N=36$, $\alpha_1=\alpha_2=2$, $\epsilon=2$, $d_{S\!R}=30~{\rm m}$, $d_{R\!D}=40~{\rm m}$, $r_e=200~{\rm m}$, and $\lambda_e=10^{-3}$. We see from Fig.~\ref{fig2} that both the analytical CDF of the received SNRs at the legitimate user $D$ and the eavesdroppers $E$ characterized by (\ref{equal10}) and (\ref{equal16}), respectively, match well with the numerical curves. In addition, the asymptotic CDF of the received SNR at the eavesdroppers $E$ calculated from (\ref{equal110}) for large $r_e$ is also verified to be quite accurate.

\subsection{Secrecy Outage Probability}
\begin{figure}[!t]
    \setlength{\abovecaptionskip}{0pt}
    \setlength{\belowcaptionskip}{0pt}
    \centering
    \includegraphics[width=3.5in]{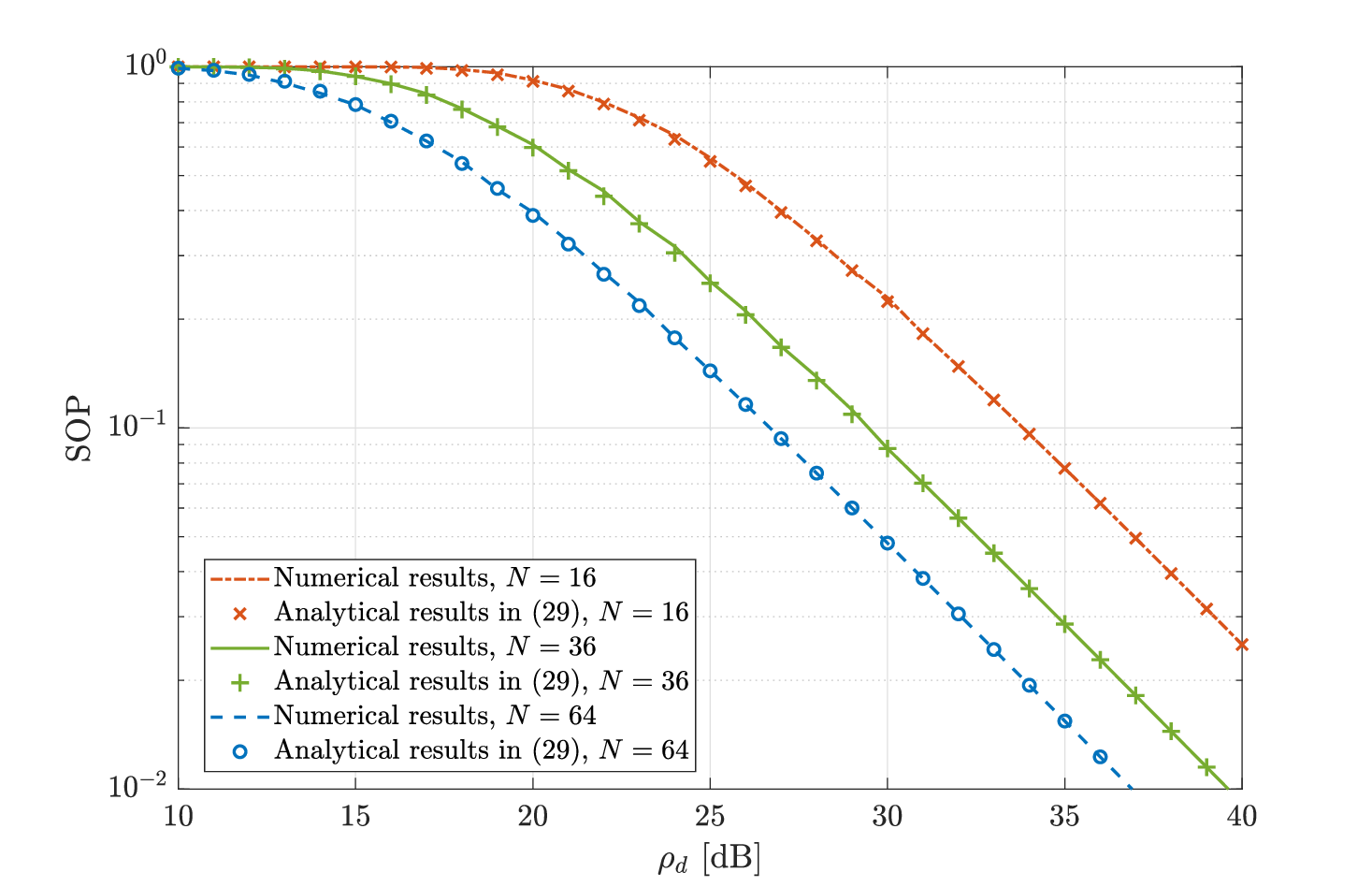}
    \caption{The SOP versus $\rho_d$, with $K=16$, $\alpha_1=\alpha_2=2$, $\epsilon=2$, $d_{S\!R}=30~{\rm m}$, $d_{R\!D}=40~{\rm m}$, $r_e=200~{\rm m}$, $\lambda_e=10^{-3}$, $C_{\rm th}=0.05$, and $\rho_e= 30~{\rm dB}$.}
    \label{fig3} \end{figure}

\begin{figure}[!t]
    \setlength{\abovecaptionskip}{0pt}
    \setlength{\belowcaptionskip}{0pt}
    \centering
    \includegraphics[width=3.5in]{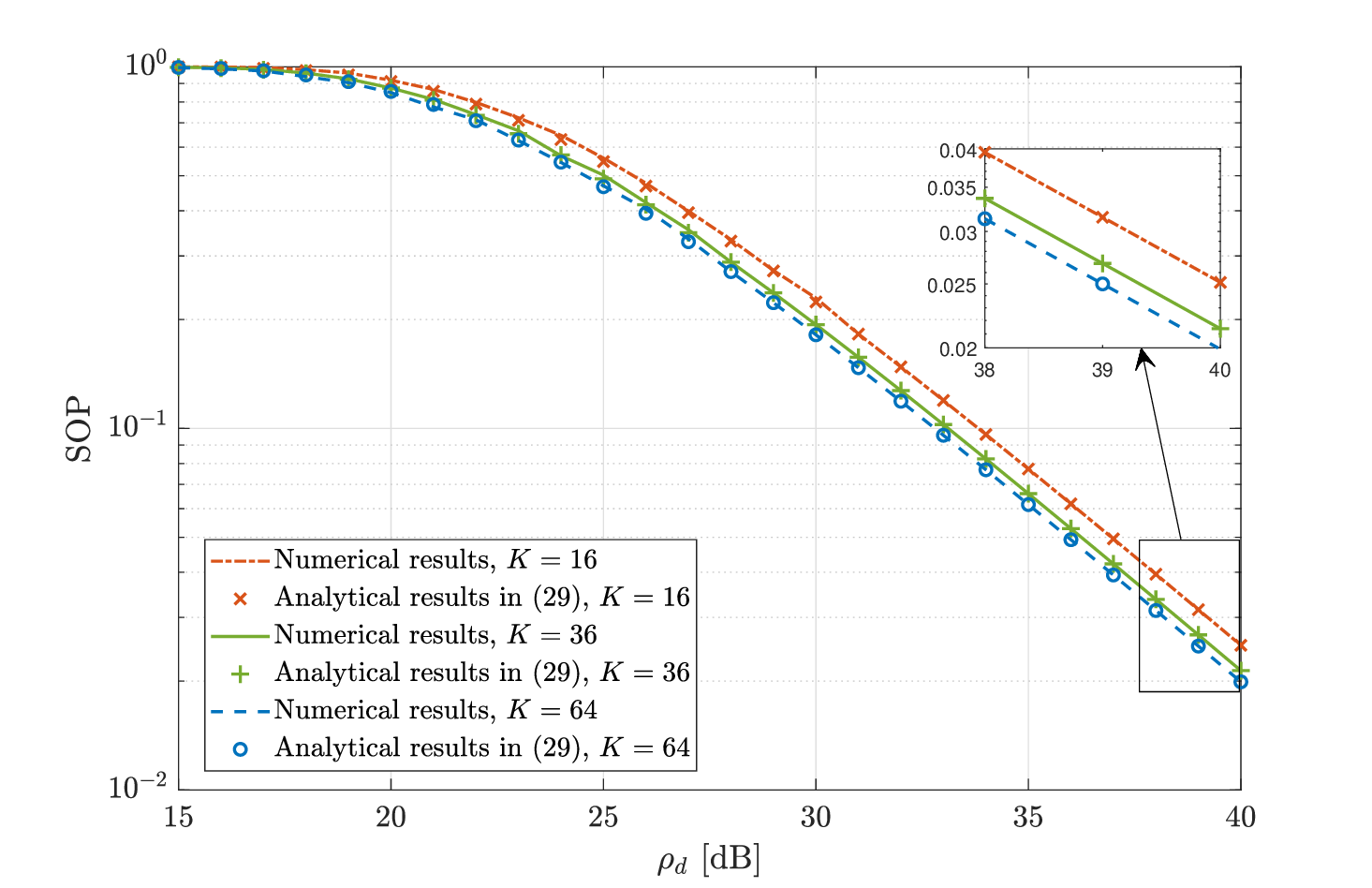}
    \caption{The SOP versus $\rho_d$, with $N=16$, $\alpha_1=\alpha_2=2$, $\epsilon=2$, $d_{S\!R}=30~{\rm m}$, $d_{R\!D}=40~{\rm m}$, $r_e=200~{\rm m}$, $\lambda_e=10^{-3}$, $C_{\rm th}=0.05$, and $\rho_e= 30~{\rm dB}$.}
    \label{fig4} \end{figure}

In this subsection, we compare the SOP obtained from Monte-Carlo simulations and the analytical results calculated from (\ref{equal22}).
Fig.~\ref{fig3} and Fig.~\ref{fig4} plot the SOP versus the transmit SNR $\rho_d$ for different values of $N$ and $K$, respectively. Again, the analytical expressions in (\ref{equal22}) match very well with the numerical results, and the SOP always decreases as the transmit SNR $\rho_d$ increases. Furthermore, as expected from \emph{Remark~4}, the SOP obviously decreases as $N$ increases. However, the SOP remains almost the same when $K$ increases with fixed $N$, which means that the impact of the number of transmit antennas on the secrecy outage performance is negligible.

\begin{figure}[!t]
    \setlength{\abovecaptionskip}{0pt}
    \setlength{\belowcaptionskip}{0pt}
    \centering
    \includegraphics[width=3.5in]{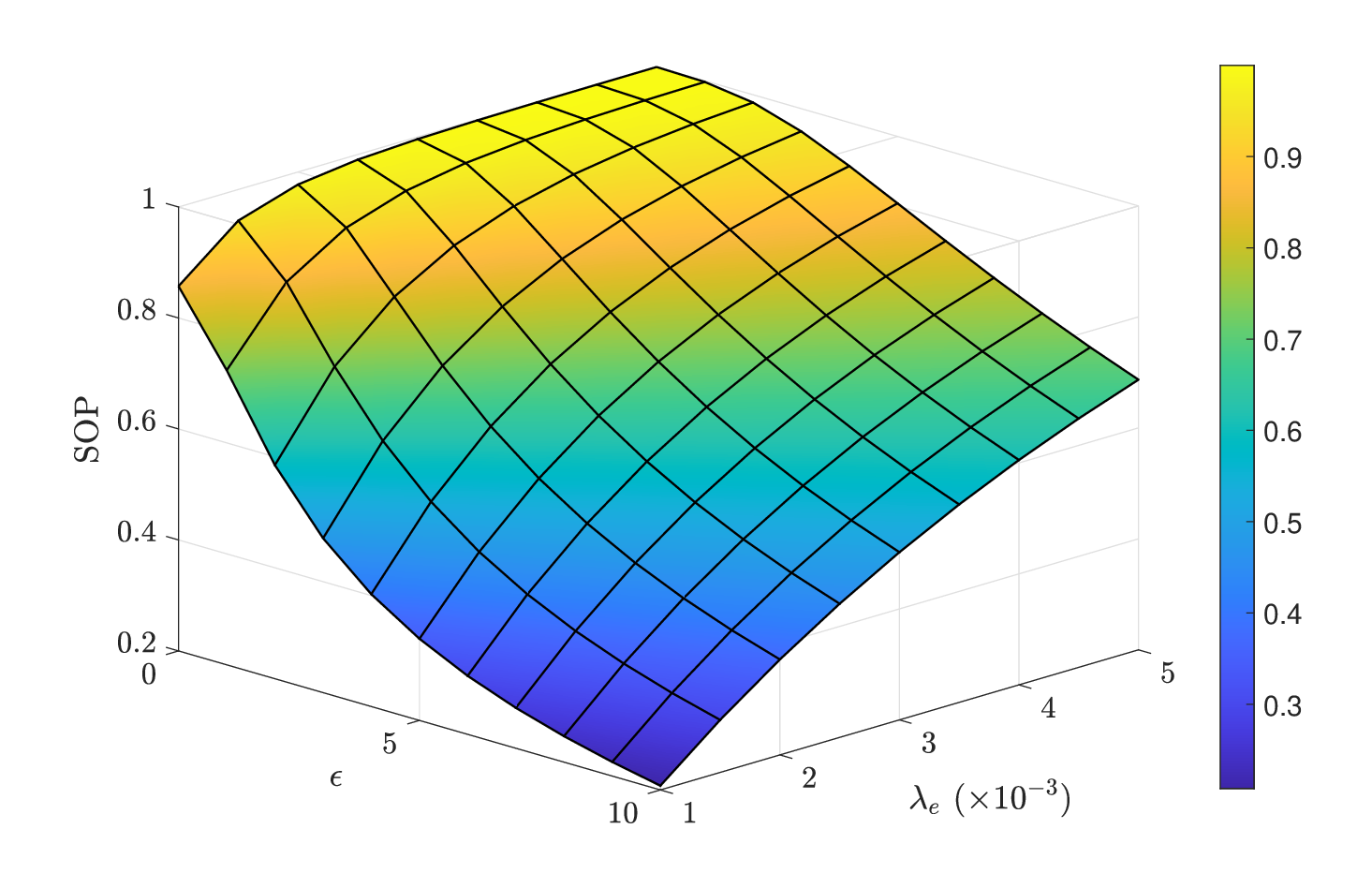}
    \caption{The SOP versus $\epsilon$ and $\lambda_e$, with $K=16$, $N=36$, $\alpha_1=\alpha_2=2$, $d_{S\!R}=30~{\rm m}$, $d_{R\!D}=40~{\rm m}$, $r_e=200~{\rm m}$, $C_{\rm th}=0.05$, $\rho_d= 20~{\rm dB}$, and $\rho_e= 30~{\rm dB}$.}
    \label{fig5} \end{figure}

Fig.~\ref{fig5} depicts the SOP as a function of the Rician factor $\epsilon$ and the eavesdropper density $\lambda_e$. It is observed that the SOP improves as $\epsilon$ increases. This is because with a large Rician factor, the channels are dominated by the LoS component with better link quality than NLoS. Additionally, it is noteworthy that the SOP increases with $\lambda_e$, since a larger $\lambda_e$ leads to more eavesdroppers which increases the likelihood that the worst-case eavesdropper obtains higher quality information.

\begin{figure}[!t]
    \setlength{\abovecaptionskip}{0pt}
    \setlength{\belowcaptionskip}{0pt}
    \centering
    \includegraphics[width=3.5in]{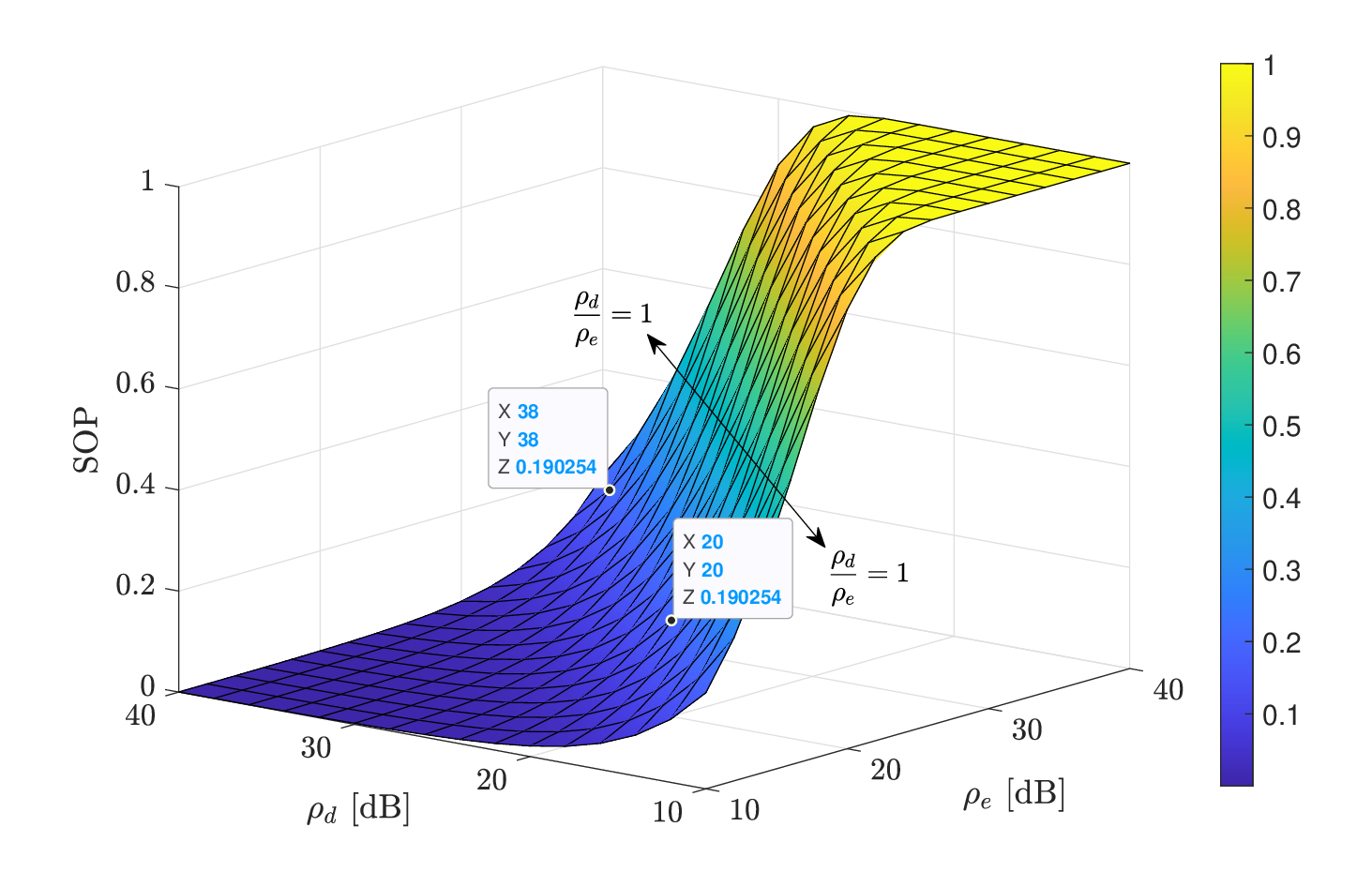}
    \caption{{The SOP versus $\rho_d$ and $\rho_e$, with $K=16$, $N=16$, $\alpha_1=\alpha_2=2$, $\epsilon=2$, $d_{S\!R}=30~{\rm m}$, $d_{R\!D}=40~{\rm m}$, $r_e=200~{\rm m}$, $\lambda_e=10^{-3}$, and $C_{\rm th}=0.05$.}}
    \label{fig6} \end{figure}

Fig.~\ref{fig6} illustrates the SOP as a function of the transmit SNRs $\rho_d$ and $\rho_e$. It is clearly shown that increasing $\rho_d$ and decreasing $\rho_e$ improve the SOP performance. We can further observe that the SOP remains the same when the ratio of the transmit SNRs at the legitimate user and the eavesdroppers, i.e., $\frac{\rho_d}{\rho_e}$, is kept fixed. This implies that increasing the transmit power $P_T$ will not help improve the secrecy outage performance of the system, which is consistent with \emph{Remark~3}.

\begin{figure}[!t]
    \setlength{\abovecaptionskip}{0pt}
    \setlength{\belowcaptionskip}{0pt}
    \centering
    \includegraphics[width=3.5in]{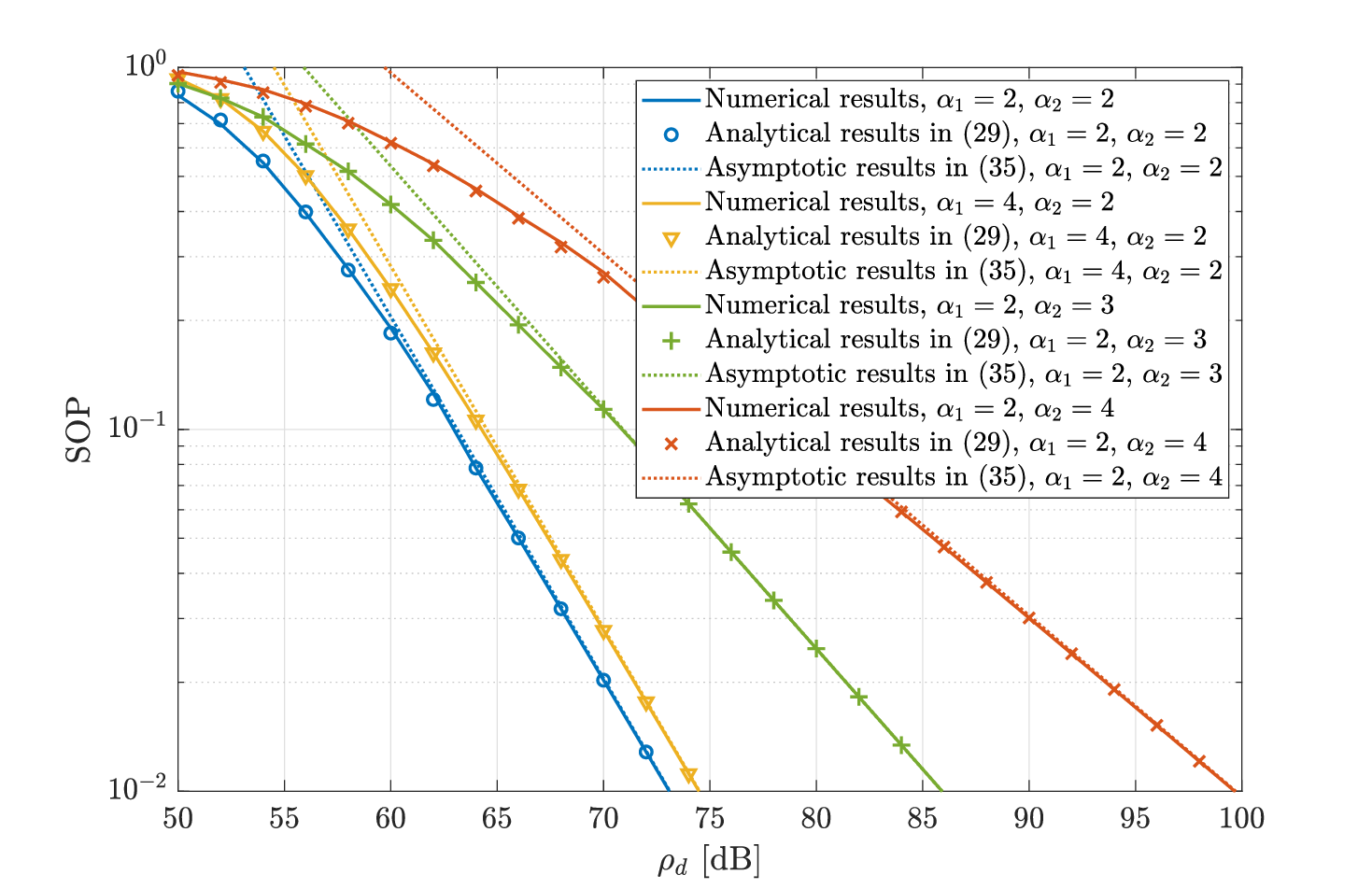}
    \caption{{The SOP versus $\rho_d$, with $K=16$, $N=16$, $\epsilon=2$, $d_{S\!R}=30~{\rm m}$, $d_{R\!D}=40~{\rm m}$, $r_e=200~{\rm m}$, $\lambda_e=10^{-3}$, $C_{\rm th}=0.05$, and $\rho_e= 60~{\rm dB}$.}}
    \label{fig7} \end{figure}

{Fig.~\ref{fig7} shows the SOP versus the transmit SNR $\rho_d$ for various values of the path loss exponents $\alpha_1$ and $\alpha_2$. It can be seen that the slope of the secrecy outage curves is determined by $\alpha_2$, which becomes less steep as $\alpha_2$ increases. Besides, according to the definition in (\ref{equal24}), the secrecy diversity order presented in \emph{Remark~5} can be demonstrated by calculating the negative slope of the SOP curves on a log-log scale.}

\subsection{Ergodic Secrecy Capacity}
\begin{figure}[!t]
    \setlength{\abovecaptionskip}{0pt}
    \setlength{\belowcaptionskip}{0pt}
    \centering
    \includegraphics[width=3.5in]{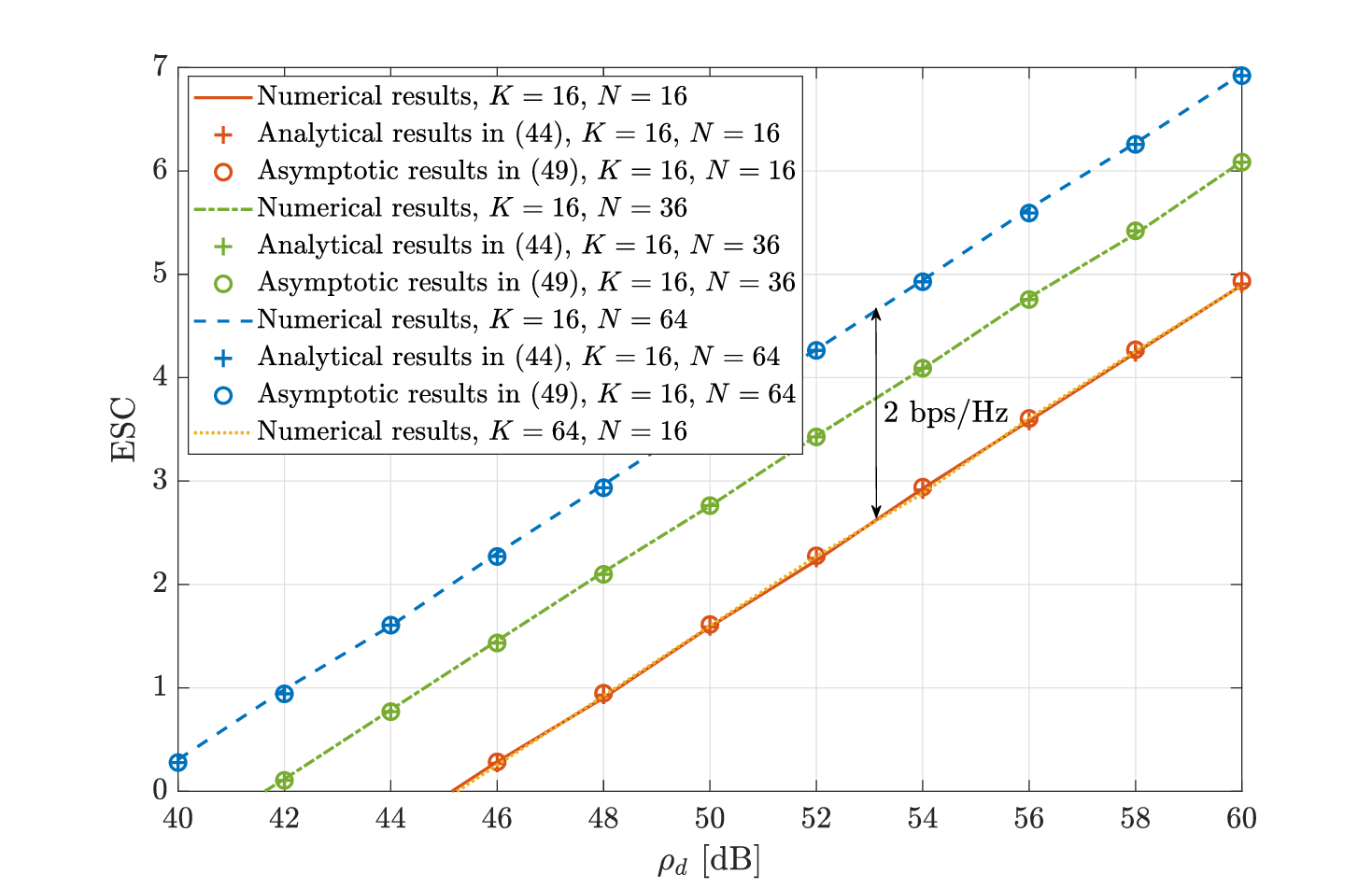}
    \caption{{The ESC versus $\rho_d$, with $\alpha_1=\alpha_2=2$, $\epsilon=2$, $d_{S\!R}=30~{\rm m}$, $d_{R\!D}=40~{\rm m}$, $r_e=200~{\rm m}$, $\lambda_e=10^{-3}$, and $\rho_e= 50~{\rm dB}$.}}
    \label{fig8} \end{figure}

In this subsection, we compare the ESC obtained from Monte-Carlo simulations with the analytical and asymptotic results calculated from (\ref{equal37}) and (\ref{equal46}), respectively.
Fig.~\ref{fig8} illustrates the ESC versus the transmit SNR $\rho_d$ for different values of $N$ and $K$. We can see that both the analytical and asymptotic expressions match well with the numerical results, and the ESC increases with $\rho_d$. As expected from \emph{Remark~6}, we can observe that the ESC at high SNR is independent of the number of transmit antennas, $K$, but obviously increases with the number of RIS reflecting elements, $N$. The ESC increases by approximately $2$ bps/Hz when the number of RIS reflecting elements increases by a factor of $4$, validating the conclusions presented in \emph{Remark~6}, i.e., the ESC increases logarithmically with the number of RIS reflecting elements.

\begin{figure}[!t]
    \setlength{\abovecaptionskip}{0pt}
    \setlength{\belowcaptionskip}{0pt}
    \centering
    \includegraphics[width=3.5in]{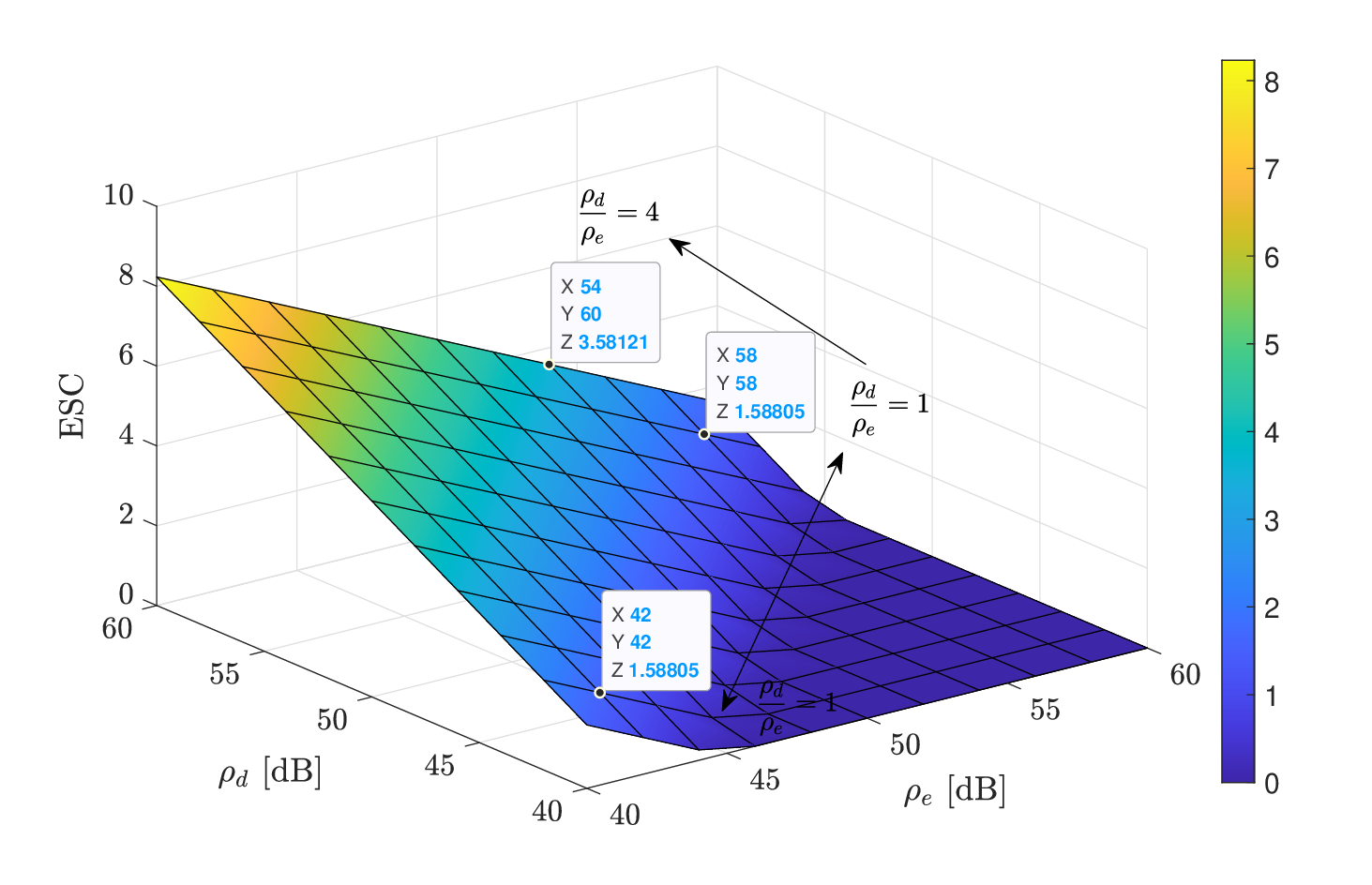}
    \caption{{The ESC versus $\rho_d$ and $\rho_e$, with $K=16$, $N=16$, $\alpha_1=\alpha_2=2$, $\epsilon=2$, $d_{S\!R}=30~{\rm m}$, $d_{R\!D}=40~{\rm m}$, $r_e=200~{\rm m}$, and $\lambda_e=10^{-3}$.}}
    \label{fig9} \end{figure}

Fig.~\ref{fig9} plots the ESC as a function of the transmit SNRs $\rho_d$ and $\rho_e$. First, similar to the results shown in Fig.~\ref{fig6}, we see that increasing $\rho_d$ and decreasing $\rho_e$ both improve the secrecy performance. Also, as predicted in \emph{Remark~8}, the ESC remains constant for a fixed ratio $\frac{\rho_d}{\rho_e}$, regardless of the specific value of the transmit power $P_T$. In addition, in Fig.~\ref{fig9}, when the ratio $\frac{\rho_d}{\rho_e}$ increases by a factor of $4$, the ESC increases by $2$ bps/Hz, which confirms the results derived in (\ref{equal46}).

\begin{figure}[!t]
    \setlength{\abovecaptionskip}{0pt}
    \setlength{\belowcaptionskip}{0pt}
    \centering
    \includegraphics[width=3.5in]{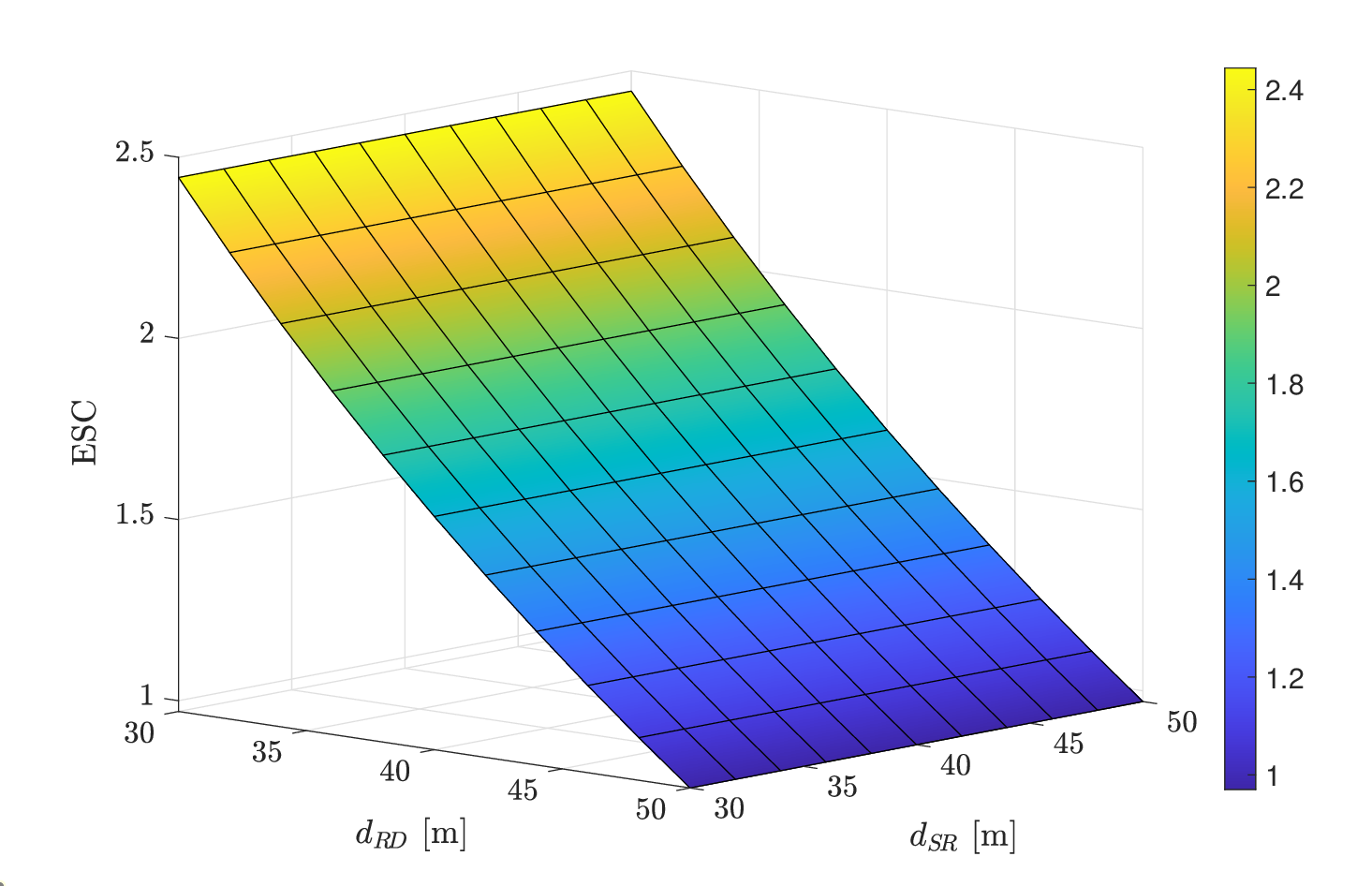}
    \caption{The ESC versus $d_{S\!R}$ and $d_{R\!D}$, with $K=16$, $N=16$, $\alpha_1=\alpha_2=2$, $\epsilon=2$, $r_e=200~{\rm m}$, $\lambda_e=10^{-3}$, $\rho_d= 50~{\rm dB}$, and $\rho_e= 50~{\rm dB}$.}
    \label{fig10} \end{figure}

Fig.~\ref{fig10} shows the ESC as a function of the distances $d_{S\!R}$ and $d_{R\!D}$. It is clear that the ESC improves as $d_{R\!D}$ decreases, and it keeps unchanged with arbitrary variation of $d_{S\!R}$, which has been validated by \emph{Remark~7}. This phenomenon can be explained by the fact that decreasing the distance from the base station to the RIS will simultaneously increase the received SNRs for both the legitimate user and the eavesdroppers.

\begin{figure}[!t]
    \setlength{\abovecaptionskip}{0pt}
    \setlength{\belowcaptionskip}{0pt}
    \centering
    \includegraphics[width=3.5in]{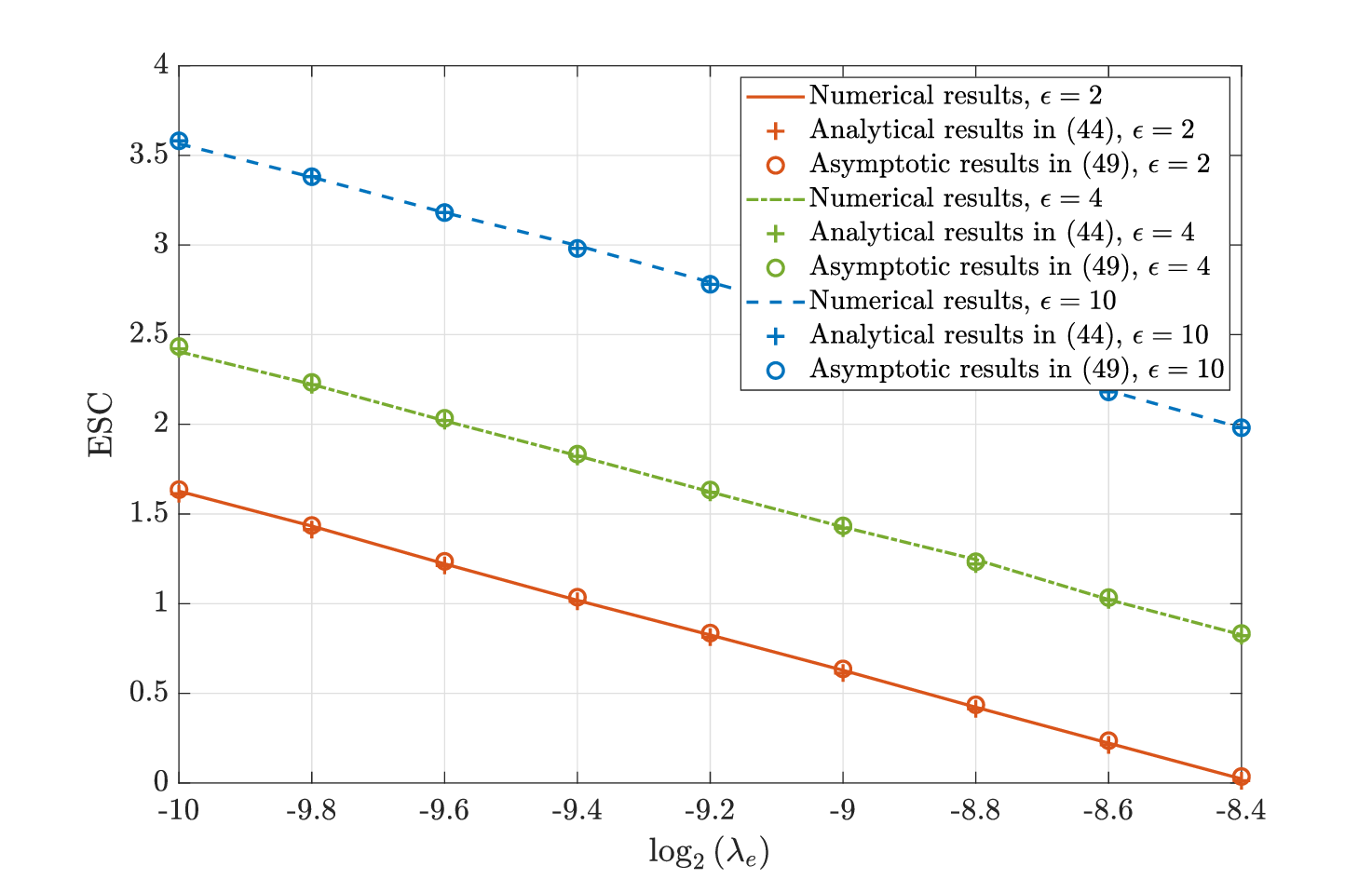}
    \caption{{The ESC versus $\log_2{\left({\lambda_e}\right)}$, with $K=16$, $N=16$, $\alpha_1=\alpha_2=2$, $d_{S\!R}=30~{\rm m}$, $d_{R\!D}=40~{\rm m}$, $r_e=200~{\rm m}$, $\rho_d= 50~{\rm dB}$, and $\rho_e= 50~{\rm dB}$.}}
    \label{fig11} \end{figure}

{Fig.~\ref{fig11} presents the ESC versus the density $\lambda_e$ of the randomly located eavesdroppers in the logarithmic domain. It is observed that the ESC decreases linearly with $\log_2{\left({\lambda_e}\right)}$. This behavior is due to the fact that a larger $\lambda_e$ results in the presence of more eavesdroppers within a fixed range, which degrades the secrecy performance. In addition, it can also be seen from Fig.~\ref{fig11} that the ESC increases with the Rician factor $\epsilon$. This is because the channels are mainly influenced by the LoS component when the Rician factor is large.}
{\subsection{Cases for A Relatively Small $r_e$ and Rician Fading Model}}
{In this subsection, we verify the generality of the analytical results of the SOP and ESC with a relatively small value of $r_e$, i.e., $r_e=50~{\rm m}$, and under the assumption that the S-RIS and RIS-$i$ channels both follow Rician fading, expressed as}
{\begin{align}
\mathbf{H}_{S\!R}=\sqrt\nu \left(\sqrt{\frac{\epsilon_1}{\epsilon_1+1}}{\overline{\mathbf{H}}}_{S\!R}+\sqrt{\frac{1}{\epsilon_1+1}}{\widetilde{\mathbf{H}}}_{S\!R}\right),
\label{R7}
\end{align}}
{\begin{align}
\mathbf{h}_{Ri}=\sqrt{\mu_i}\left(\sqrt{\frac{\epsilon_2}{\epsilon_2+1}}{\overline{\mathbf{h}}}_{Ri}+\sqrt{\frac{1}{\epsilon_2+1}}{\widetilde{\mathbf{h}}}_{Ri}\right),
\label{R8}
\end{align}}
{where $\epsilon_1$ and $\epsilon_2$ are the Rician factors for the $S$-RIS and RIS-$i$ channels, respectively. Note that when the Rician factor $\epsilon_1$ is large, (\ref{R7}) simplifies to the LoS model we consider in~(\ref{equal1}). Therefore, the Rician factor $\epsilon_1$ is set to be small, e.g., $\epsilon_1=2$.}

\begin{figure}[!t]
    \setlength{\abovecaptionskip}{0pt}
    \setlength{\belowcaptionskip}{0pt}
    \centering
    \includegraphics[width=3.5in]{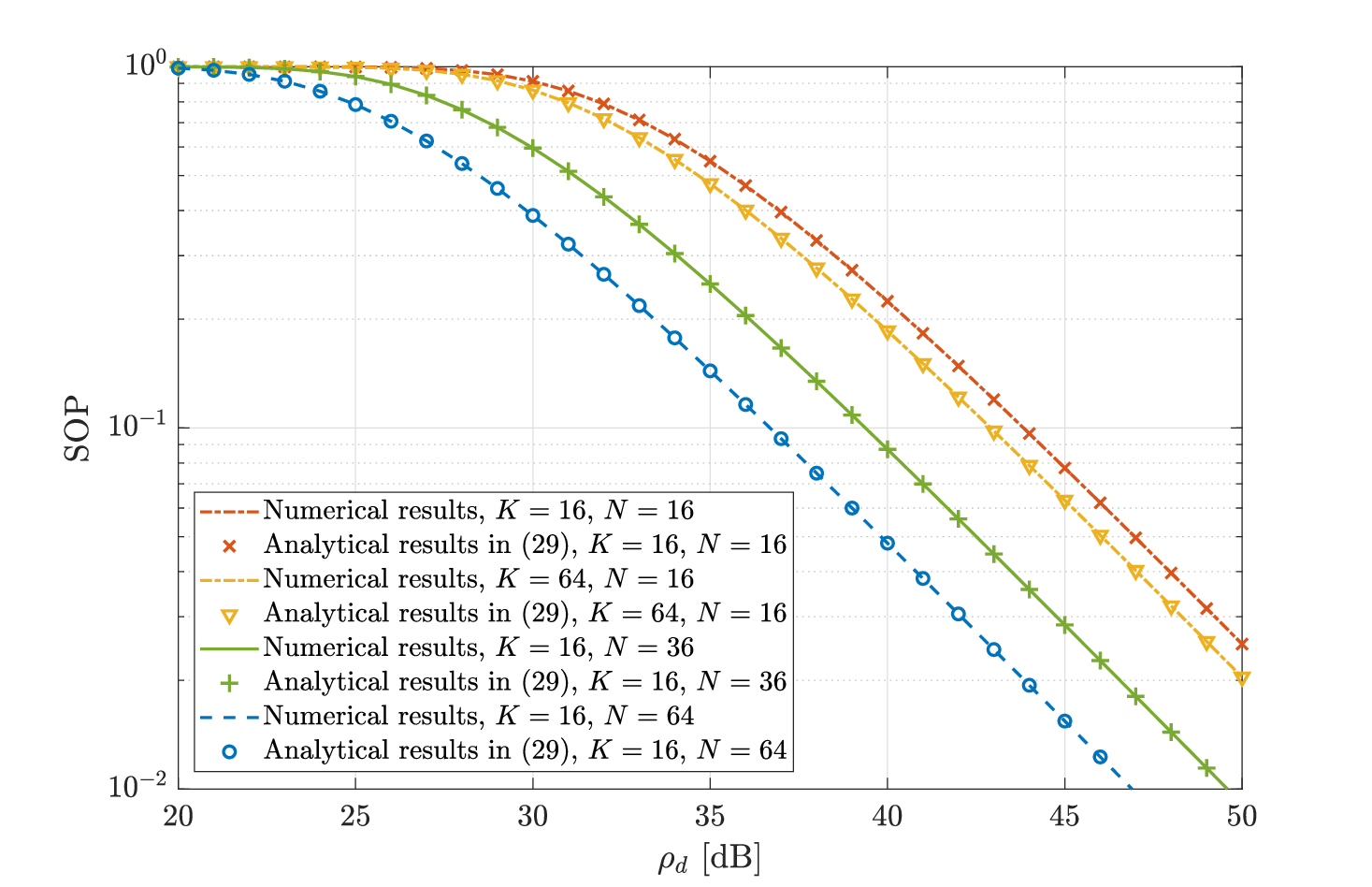}
    \caption{{The SOP versus $\rho_d$, with $\alpha_1\!=\!\alpha_2\!=\!2$, $\epsilon_1\!=\!\epsilon_2\!=\!2$, $d_{S\!R}\!=\!30~{\rm m}$, $d_{R\!D}\!=\!40~{\rm m}$, $r_e\!=\!50~{\rm m}$, $\lambda_e\!=\!10^{-2}$, $C_{\rm th}=0.05$, and $\rho_e\!=\! 30~{\rm dB}$.}}
    \vspace{-0.45cm}
    \label{response5} \end{figure}

\begin{figure}[!t]
    \setlength{\abovecaptionskip}{0pt}
    \setlength{\belowcaptionskip}{0pt}
    \centering
    \includegraphics[width=3.5in]{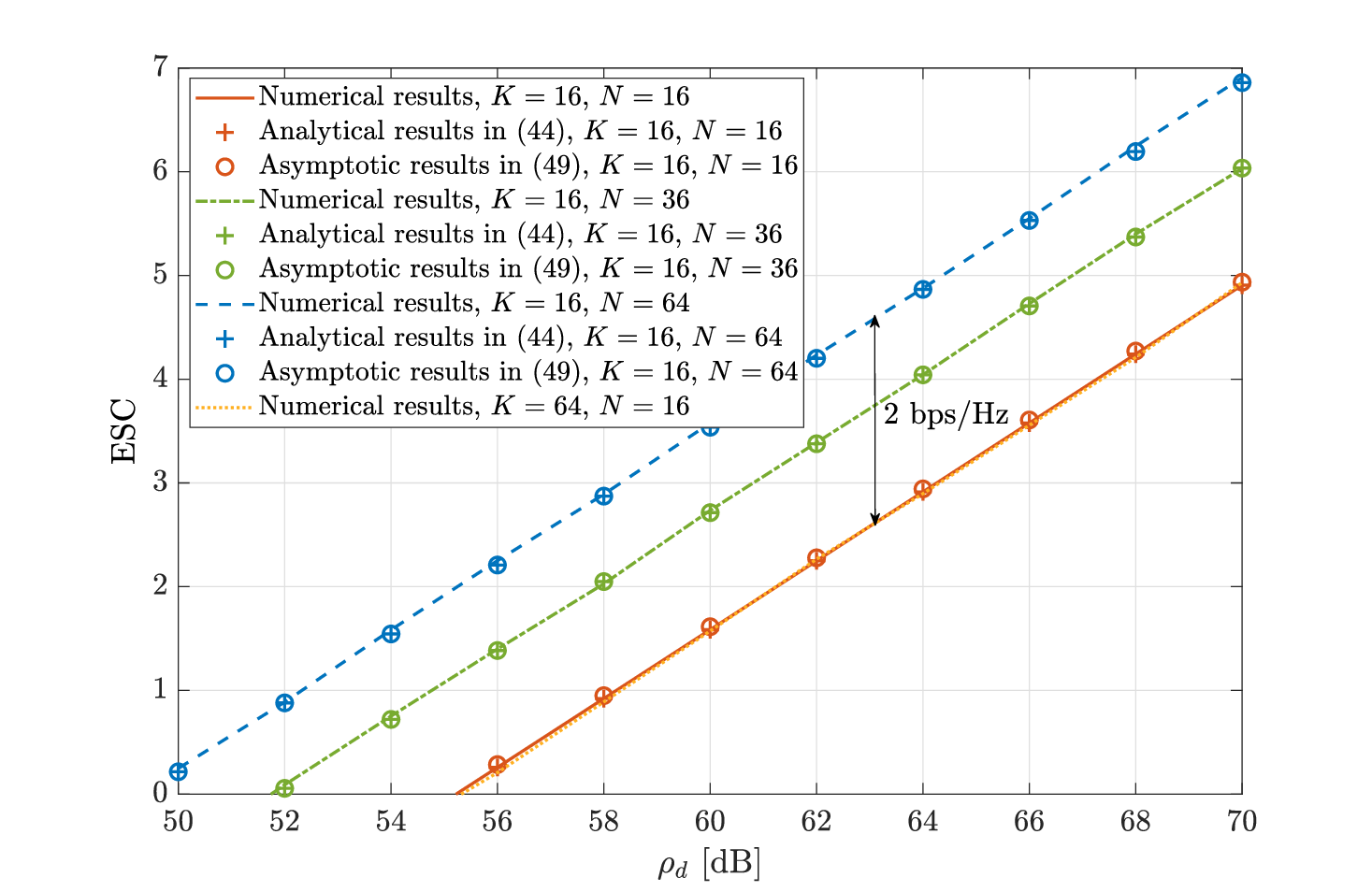}
    \caption{{The ESC versus $\rho_d$, with $\alpha_1\!=\!\alpha_2\!=\!2$, $\epsilon_1\!=\!\epsilon_2\!=\!2$, $d_{S\!R}\!=\!30~{\rm m}$, $d_{R\!D}\!=\!40~{\rm m}$, $r_e\!=\!50~{\rm m}$, $\lambda_e\!=\!10^{-2}$, and $\rho_e\!=\!50~{\rm dB}$.}}
    \label{response7} \end{figure}

{As shown in Fig.~\ref{response5} and Fig.~\ref{response7}, it can be seen that the numerical results with small $r_e$ and Rician fading model still match well with the analytical results under the assumption of large $r_e$ and LoS model. Also, the observations that the SOP and ESC are mainly affected by the number of RIS reflecting elements rather than the number of transmit antennas, which is obtained with large $r_e$ and LoS model, are also applicable.}

\section{Conclusion}
In this paper, the secrecy performance of an RIS-assisted communication system with spatially random eavesdroppers was studied. Specifically, the exact distributions of the received SNRs at the legitimate user and the eavesdroppers were first derived. Then, based on these distributions, the closed-form SOP and ESC expressions were derived. Furthermore, a high-SNR secrecy diversity order and the asymptotic ESC analysis have also been conducted. The obtained results quantify the impact of key parameters on the secrecy performance and provide insightful guidelines for system design. Several simulations were provided to demonstrate the obtained results.
\begin{appendices}
\section{Proof of Theorem 1}
For the transmit beamformer $\mathbf{f}=\frac{\mathbf{g}_D}{\left\|\mathbf{g}_D^H\right\|}$, we compute the optimal reflecting phase shifts at the RIS by maximizing the received signal power as follows
{\begin{align}
\mathbf{\Theta}^{\rm opt}=\arg\mathop{\max}\limits_{\mathbf{\Theta}}&{\left|\mathbf{g}_D^H\mathbf{f}\right|^2}=\arg\mathop{\max}\limits_{\mathbf{\Theta}}{\left\|\mathbf{h}_{R\!D}^H\mathbf{\Theta}\mathbf{H}_{S\!R}\right\|^2}\nonumber\\
=\arg\mathop{\max}\limits_{\mathbf{\Theta}}&{\left|\mathbf{h}_{R\!D}^H\mathbf{\Theta}\mathbf{a}_{N,S\!R}\right|^2\left\|\mathbf{a}_{K,S\!R}^H \right\|^2}\nonumber\\
\mathop=^{\left(\rm d\right)}\arg\mathop{\max}\limits_{\mathbf{\Theta}}&{\left|{\bm{\theta}}^H{\rm diag}\left\{\mathbf{h}_{R\!D}^H\right\}\mathbf{a}_{N,S\!R}\right|^2}\nonumber\\
=\arg\mathop{\max}\limits_{\mathbf{\Theta}}&\left|\sum_{n=1}^{N}h_{R\!D}^\ast\left(n\right){\rm a}_{N,S\!R}\left(n\right){\rm e}^{j\theta_n}\right|^2,
\label{equal100}
\end{align}}
\hspace*{-0.2cm} where $({\rm d})$ follows by defining $\bm{\theta}^H\!=\!\!\left[{\rm e}^{j\theta_1},\ldots,{\rm e}^{j\theta_n},\ldots,{\rm e}^{j\theta_N}\right]$, and using the identity $\left\|\mathbf{a}_{K,S\!R}^H \right\|^2=K$. {From (\ref{equal100}), we see that maximizing $\left|\mathbf{g}_D^H\mathbf{f}\right|^2$ is equivalent to ensuring the phases of $N$ complex RVs $h_{R\!D}^\ast\left(n\right){\rm a}_{N,S\!R}\left(n\right){\rm e}^{j\theta_n}$ being identical.} Therefore, the optimal RIS phase shifts are given by
\begin{align}
\theta_n^{\rm opt}=-\angle\left(h_{R\!D}^\ast\left(n\right){\rm a}_{N,S\!R}\left(n\right)\right),
\label{equal101}
\end{align}
and the corresponding reflection matrix can be easily obtained as (\ref{equal8}). 
This completes the proof.

\section{Proof of Lemma 1}
Since $\left|h_{R\!D}\left(1\right)\right|,\left|h_{R\!D}\left(2\right)\right|,\ldots,\left|h_{R\!D}\left(N\right)\right|$ are i.i.d. RVs, the mean and variance of the RV $\left|{\rm A}\right|=\sqrt{K\nu}\sum_{n=1}^{N}\left|h_{R\!D}\left(n\right)\right|$ can be respectively calculated as
\begin{equation}
\mathbb{E}\{\left|{\rm A}\right|\}=\sqrt{K\nu}N\cdot\mathbb{E}\{\left|h_{R\!D}\left(n\right)\right|\},
\label{equal103}
\end{equation}
and
\begin{equation}
{\rm Var}\{\left|{\rm A}\right|\}=K\nu N\cdot{\rm Var}\{\left|h_{R\!D}\left(n\right)\right|\},
\label{equal104}
\end{equation}
where $\left|h_{R\!D}\left(n\right)\right|\!\sim\! Rice\left(\sqrt{\frac{\mu_D\epsilon}{\epsilon+1}},\sqrt{\frac{1}{2}\frac{\mu_D}{\epsilon+1}}\right)$, whose mean and variance are given as $\mathbb{E}\{\left|h_{R\!D}\left(n\right)\right|\}\!\!=\!\!\sqrt{\frac{\mu_D}{\epsilon+1}}\frac{\sqrt\pi}{2}L_\frac{1}{2}\left(-\epsilon\right)$ and ${\rm Var}\{\left|h_{R\!D}\left(n\right)\right|\}=\frac{\mu_D}{\epsilon+1}\left[1+\epsilon-\frac{\pi}{4}\left(L_\frac{1}{2}\left(-\epsilon\right)\right)^2\right]$, respectively. 
Therefore, according to \cite{18}, the RV $\left|{\rm A}\right|$ can be approximated by a Gamma distributed RV with shape parameter $k=\frac{\mathbb{E}\{\left|{\rm A}\right|\}^2}{{\rm Var}\{\left|{\rm A}\right|\}}$ and scale parameter $\theta=\frac{{\rm Var}\{\left|{\rm A}\right|\}}{\mathbb{E}\{\left|{\rm A}\right|\}}$, which yields the desired result in (\ref{equal9}).

\section{Proof of Proposition 1}
The received SNR at $E_m$ can be written as
\begin{align}
\gamma_{E_m}&=\rho_e \left|\mathbf{h}_{R\!E_m}^H\mathbf{\Theta}\mathbf{H}_{S\!R}\mathbf{f}\right|^2\nonumber\\
&\mathop=^{\left(\rm e\right)}\rho_e \left|\mathbf{h}_{R\!E_m}^H\mathbf{\Theta}\mathbf{H}_{S\!R}\frac{\mathbf{H}_{S\!R}^H\mathbf{\Theta}^H\mathbf{h}_{R\!D}}{\left\|\mathbf{h}_{R\!D}^H\mathbf{\Theta}\mathbf{H}_{S\!R}\right\|}\right|^2\nonumber\\
&=\rho_e\nu\frac{\left|\mathbf{h}_{RE_m}^H\mathbf{\Theta}\mathbf{a}_{N,SR}\right|^2\left|\mathbf{a}_{N,SR}^H\mathbf{\Theta}^H\mathbf{h}_{RD}\right|^2\left\|\mathbf{a}_{K,SR}^H\right\|^4}{\left|\mathbf{h}_{RD}^H\mathbf{\Theta}\mathbf{a}_{N,SR}\right|^2\left\|\mathbf{a}_{K,SR}^H\right\|^2}\nonumber\\
&=\rho_eK\nu\left|\mathbf{h}_{RE_m}^H\mathbf{\Theta}\mathbf{a}_{N,SR}\right|^2\nonumber\\
&=\rho_eK\nu\left|\sum_{n=1}^{N}{h_{RE_m}^\ast\left(n\right){\rm a}_{N,SR}\left(n\right){\rm e}^{j\theta_n}}\right|^2\nonumber\\
&\mathop=^{\left(\rm f\right)}\rho_eK\nu\left|\sum_{n=1}^{N}{h_{RE_m}^\ast\left(n\right){\rm e}^{-j\angle h_{RD}^\ast\left(n\right)}}\right|^2,
\label{equal105}
\end{align}
where $({\rm e})$ is obtained by substituting the expressions of $\mathbf{f}$ and $\mathbf{g}_D^H$, and $({\rm f})$ comes from (\ref{equal101}) with $\left|{\rm a}_{N,SR}\left(n\right)\right|=1$.

\section{Proof of Lemma 2}
From (\ref{equal2}), we see that $h_{Ri}\left(n\right)\!\sim\!\mathcal{CN}\left(\sqrt{\frac{\mu_i\epsilon}{\epsilon+1}}{\bar{h}}_{Ri}\left(n\right),\frac{\mu_i}{\epsilon+1}\right)$, and $\left|h_{Ri}\left(n\right)\right|\!\sim\! Rice\left(\sqrt{\frac{\mu_i\epsilon}{\epsilon+1}},\sqrt{\frac{1}{2}\frac{\mu_i}{\epsilon+1}}\right)$. Then, it can be easily obtained that $\mathbb{E}\{h_{Ri}\!\left(n\right)\}\!=\!\sqrt{\frac{\mu_i\epsilon}{\epsilon+1}}{\rm a}_{N,Ri}\!\left(n\right)$, $\mathbb{E}\{\left|h_{Ri}\!\left(n\right)\right|\}\!=\!\sqrt{\frac{\mu_i}{\epsilon+1}}\frac{\sqrt\pi}{2}L_\frac{1}{2}\!\left(-\epsilon\right)$, and $\mathbb{E}\{\left|h_{Ri}\left(n\right)\right|^2\}\!=\!\mu_i$. It follows that
\begin{align}
\mathbb{E}\{{\rm e}^{-j\angle h_{RD}^\ast\left(n\right)}\}\!=\!\left(\frac{\mathbb{E}\{h_{RD}^\ast\left(n\right)\}}{\mathbb{E}\{\left|h_{RD}^\ast\left(n\right)\right|\}}\right)^\ast\!=\!\frac{\sqrt\epsilon {\rm a}_{N,RD}\left(n\right)}{\frac{\sqrt\pi}{2}L_\frac{1}{2}\left(-\epsilon\right)}.
\label{equal106}
\end{align}

Therefore, the mean and variance of the RV $x_n=h_{RE_m}^\ast\left(n\right){\rm e}^{-j\angle h_{RD}^\ast\left(n\right)}$ can be calculated, respectively, as
\begin{align}
\mathbb{E}\{x_n\}=\sqrt{\frac{\mu_{E_m}\epsilon^2}{\epsilon+1}}\frac{{\rm a}_{N,RE_m}^\ast\left(n\right){\rm a}_{N,RD}\left(n\right)}{\frac{\sqrt\pi}{2}L_\frac{1}{2}\left(-\epsilon\right)},
\label{equal107}
\end{align}
and
\begin{align}
{\rm Var}\{x_n\}=\mu_{E_m}\left[1-\frac{\epsilon^2}{{\frac{\pi}{4}\left(\epsilon+1\right)\left(L_\frac{1}{2}\left(-\epsilon\right)\right)}^2}\right].
\label{equal108}
\end{align}

It can be seen from (\ref{equal107}) that $\mathbb{E}\{x_n\}$ depends on $n$, which means that $x_n$ is not identically distributed. Therefore, the distribution of the RV $Z_{E_m}=\sum_{n=1}^{N}x_n$ cannot be directly approximated as Gaussian by applying the central limit theorem (CLT). In order to characterize the distribution of $Z_{E_m}$, we first define a new RV $x_n-\mathbb{E}\{x_n\}$, and it can be easily verified that $x_1-\mathbb{E}\{x_1\},x_2-\mathbb{E}\{x_2\},\ldots,x_N-\mathbb{E}\{x_N\}$ are i.i.d. RVs with zero mean and variance ${\rm Var}\{x_n\}$. By virtue of the CLT, $\sum_{n=1}^{N}\left(x_n-\mathbb{E}\{x_n\}\right)$ converges in distribution to a complex Gaussian RV with zero mean and variance $N\!\cdot\!{\rm Var}\{x_n\}$. Then, we can obtain that $Z_{E_m}=\sum_{n=1}^{N}x_n\!=\!\sum_{n=1}^{N}\left(x_n-\mathbb{E}\{x_n\}\right)+\sum_{n=1}^{N}\mathbb{E}\{x_n\}\sim\mathcal{CN}\left(\sum_{n=1}^{N}\mathbb{E}\{x_n\},N\!\cdot\!{\rm Var}\{x_n\}\right)$, where
\begin{align}
\sum_{n=1}^{N}\mathbb{E}\{x_n\}\!=\!&\sqrt{\frac{\mu_{E_m}\epsilon^2}{{\frac{\pi}{4}\!\left(\epsilon\!+\!1\right)\!\left(\!L_\frac{1}{2}\!\left(-\epsilon\right)\right)}^2}}\sum_{n=1}^{N}{{\rm a}_{N,RE_m}^\ast\!\!\left(n\right){\rm a}_{N,RD}\!\left(n\right)}\nonumber\\
\mathop=^{\left(\rm g\right)}&\sqrt{\frac{\mu_{E_m}\epsilon^2}{{\frac{\pi}{4}\!\left(\epsilon\!+\!1\right)\!\left(\!L_\frac{1}{2}\!\left(-\epsilon\right)\right)}^2}}\sum_{0\le x,y\le\sqrt N-1}\!\!\!\!\!\!\!\!{\rm e}^{j2\pi\frac{d}{\lambda}\left(x\delta_1+y\delta_2\right)}\nonumber\\
\!=\!&\sqrt{\frac{\mu_{E_m}\epsilon^2}{{\frac{\pi}{4}\!\left(\epsilon\!+\!1\right)\!\left(\!L_\frac{1}{2}\!\left(-\epsilon\right)\right)}^2}}{\rm e}^{j\pi\frac{d}{\lambda}\left(\sqrt N-1\right)\left(\delta_1+\delta_2\right)}\nonumber\\
&\times\frac{\sin{\left(\pi\frac{d}{\lambda}\sqrt N \delta_1\right)}\sin{\left(\pi\frac{d}{\lambda}\sqrt N \delta_2\right)}}{\sin{\left(\pi\frac{d}{\lambda}\delta_1\right)}\sin{\left(\pi\frac{d}{\lambda}\delta_2\right)}},
\label{equal109}
\end{align}
and $({\rm g})$ follows by making use of a mapping from the index $n$ to the two-dimensional index $\left(x,y\right)$ and substituting ${\rm a}_{N,Ri}\left(n\right)={\rm e}^{j2\pi\frac{d}{\lambda}\left(x\sin{\psi_{Ri}^a}\sin{\psi_{Ri}^e}+y\cos{\psi_{Ri}^e}\right)}$ from (\ref{equal5}), $\delta_1=\sin{\psi_{RD}^a}\sin{\psi_{RD}^e}-\sin{\psi_{RE_m}^a}\sin{\psi_{RE_m}^e}$, and $\delta_2=\cos{\psi_{RD}^e}-\cos{\psi_{RE_m}^e}$.

\section{Proof of Proposition 2}
When $r_e\rightarrow\infty$, the CDF of the overall eavesdropper SNR in (\ref{equal16}) is given as
\begin{align}
F_{\gamma_E}\left(x\right)=&{\rm exp}\left[-t_0x^{-t_4}\Gamma\left(t_1\right)\right]{\rm exp}\left[t_0x^{-t_4}\Gamma\left(t_1,t_2x^{t_3}\right)\right]\nonumber\\
\mathop=^{\left(\rm h\right)}&{\rm exp}\left[-t_0x^{-t_4}\Gamma\left(t_1\right)\right]\nonumber\\
&\times\left(1+t_0x^{-t_4}\mathcal{O}\left(\left(t_2x^{t_3}\right)^{t_1-1}{\rm e}^{-t_2x^{t_3}}\right)\right),
\label{equal110}
\end{align}
where $({\rm h})$ comes from the asymptotic expansion of the upper incomplete Gamma function when $t_2={\rm e}^{v\left(\varpi\right)}\Xi^{\mu\left(\varpi\right)}{r_e}^{\frac{\alpha_2}{2}\mu\left(\varpi\right)}\rightarrow\infty$ \cite[Eq.~(6.5.32)]{19}.
Therefore, for large $r_e$, the SOP in (\ref{equal18}) is further rewritten as
\begin{align}
{\rm SOP}&=1-\int_{0}^{+\infty}{F_{\gamma_E}\left(\frac{1}{\varphi}\left(1+x\right)-1\right)f_{\gamma_D}\left(x\right)}{\rm{d}}x\nonumber\\
&\mathop\simeq^{\left(\rm i\right)}1-\int_{0}^{+\infty}{\rm exp}\left[-t_0\Gamma\left(t_1\right)\left(\frac{x}{\varphi}\right)^{-t_4}\right]\nonumber\\
&~~~~~~~~~~~~\times\frac{1}{\Gamma\left(k\right)}\frac{{\rm e}^{-\frac{\sqrt{{x}/{\rho_d}}}{\theta}}\left(\frac{\sqrt{{x}/{\rho_d}}}{\theta}\right)^k}{2x}{\rm{d}}x\nonumber\\
&=1-\frac{1}{2\Gamma\left(k\right)}\left(\sqrt{\rho_d}\theta\right)^{-k}\ I,
\label{equal111}
\end{align}
where $({\rm i})$ is obtained by substituting (\ref{equal11}) and (\ref{equal110}), $I=\int_{0}^{+\infty}{x^a {\rm exp}\left[-bx^{-c}-\upsilon\sqrt x\right]}{\rm{d}}x$, $a=\frac{k}{2}-1$, $b=t_0\Gamma\left(t_1\right)\varphi^{t_4}$, $c=t_4=\frac{2q}{p}$, and $\upsilon=\frac{1}{\sqrt{\rho_d}\theta}$.

By applying the Mellin convolution theorem \cite{20}, the Mellin transform of $I$ is given by
\begin{align}
\mathcal{M}\left[I;s\right]=\frac{2p}{2q\upsilon^{2s+2a+2}}\Gamma\left(\frac{ps}{2q}\right)\Gamma\left(2s+2a+2\right).
\label{equal112}
\end{align}
Therefore, we can calculate $I$ using the inverse transform as follows
\begin{align}
I=&\frac{p}{\pi i\upsilon^{2a+2}}\int_{u-i\infty}^{u+i\infty}{\!\Gamma\left(ps\right)\!\Gamma\left(4q\left(s\!+\!\frac{a\!+\!1}{2q}\right)\right)\!\left(\upsilon^{4q}b^p\right)^{-s}}{\rm{d}}s\nonumber\\
\mathop=^{\left(\rm j\right)}&\frac{p^\frac{1}{2}q^{2a+\frac{3}{2}}}{\upsilon^{2a+2}2^{\frac{p+4q}{2}-4a-5}\pi^{\frac{p+4q}{2}-1}}\frac{1}{2\pi i}\int_{u-i\infty}^{u+i\infty}\left(\frac{\upsilon^{4q}b^p}{p^p{256}^qq^{4q}}\right)^{-s}\nonumber\\
&\times\prod_{n=0}^{p-1}\Gamma\left(s+\frac{n}{p}\right)\prod_{n=0}^{4q-1}\Gamma\left(s+\frac{n+2a+2}{4q}\right){\rm{d}}s\nonumber\\
=&\frac{p^\frac{1}{2}q^{2a+\frac{3}{2}}}{\upsilon^{2a+2}2^{\frac{p+4q}{2}-4a-5}\pi^{\frac{p+4q}{2}-1}}G_{0,p+4q}^{p+4q,0}\left(\frac{\upsilon^{4q}b^p}{p^p{256}^qq^{4q}}\middle|\begin{matrix}-\\\Delta\\\end{matrix}\right),
\label{equal113}
\end{align}
where $({\rm j})$ follows from Gauss' multiplication formula \cite[Eq.~(6.1.20)]{19}, and the last equality is derived by applying the definition of Meijer's $G$ function. Subsequently, by substituting (\ref{equal113}) into (\ref{equal111}), the SOP is obtained as shown in (\ref{equal22}).

\section{Proof of Lemma 4}


{First, by using Jensen's inequality, we have}
{\begin{align}
\!\!\!\mathbb{E}\!\left\{\log_2\!{\left(1\!+\!\gamma_D\right)}\right\}&\!\le\!\log_2\!{\left(1\!+\!\mathbb{E}\!\left\{\gamma_D\right\}\right)}\!=\!\log_2\!{\left(1\!+\!\rho_dK\nu J\right)},
\label{equal44}
\end{align}}
\newcounter{TempEqCnt}                         
\setcounter{TempEqCnt}{\value{equation}} 
\setcounter{equation}{64}                           
\begin{figure*}[t] 
\begin{align}
{\bar{C}}_s&=\left\{\log_2{\left(1+\rho_dKN\nu\mu_D\!\left(1+\frac{\pi}{4}\left(N-1\right)\frac{\left(L_{\frac{1}{2}}\left(-\epsilon\right)\right)^2}{\epsilon+1}\!\right)\right)}\right.\nonumber\\
&\left.-\frac{1}{\ln{2}}\!\left\{\gamma+\ln{t_0\Gamma\left(t_1\right)}+{\rm exp}\left(t_0\Gamma\left(t_1\right)\right)\!\times\!\left({\rm Shi}\left(t_0\Gamma\left(t_1\right)\right)-{\rm Chi}\left(t_0\Gamma\left(t_1\right)\right)\right)\right\}\right\}^+\nonumber\\
&\mathop\to\limits^{\left(\rm k\right)}\left\{\log_2{\rho_dKN\nu\mu_D\left[1+\left(N-1\right)\frac{1}{\epsilon+1}\frac{\pi}{4}\left(L_\frac{1}{2}\left(-\epsilon\right)\right)^2\right]}-\frac{1}{\ln{2}}\left\{\gamma+\ln{t_0\Gamma\left(t_1\right)}\right\}\right\}^+\nonumber\\
&=\left\{\log_2\frac{\rho_d}{\rho_e}+\log_2{\frac{\mu_D}{\pi\lambda_e\beta_0}}+\log_2\frac{1+\left(N-1\right)\frac{1}{\epsilon+1}\frac{\pi}{4}\left(L_\frac{1}{2}\left(-\epsilon\right)\right)^2}{1-\frac{\epsilon^2}{{\frac{\pi}{4}\left(\epsilon+1\right)\left(L_\frac{1}{2}\left(-\epsilon\right)\right)}^2}}-\frac{\gamma}{\ln{2}}+C\right\}^+,
\label{equal114}
\end{align}
\hrulefill
\end{figure*}
\setcounter{equation}{\value{TempEqCnt}} 
\hspace*{-0.4cm}where $J$ is expressed as
\setcounter{equation}{63} 
\begin{align}
J&=\mathbb{E}\!\left\{\left(\sum_{n=1}^{N}\left|h_{R\!D}\left(n\right)\right|\right)^2\right\}\nonumber\\
&=\mathbb{E}\!\left\{\sum_{n=1}^{N}\left|h_{R\!D}\left(n\right)\right|^2+2\sum_{1\le i<j\le N}\left|h_{R\!D}\left(i\right)\right|\left|h_{R\!D}\left(j\right)\right|\right\}\nonumber\\
&=N\cdot\mathbb{E}\!\left\{\left|h_{R\!D}\left(n\right)\right|^2\right\}\!+N\left(N-1\right)\cdot\left(\mathbb{E}\!\left\{\left|h_{R\!D}\left(n\right)\right|\right\}\right)^2\nonumber\\
&=N\mu_D\left[1+\left(N-1\right)\frac{1}{\epsilon+1}\frac{\pi}{4}\left(L_{1/2}\left(-\epsilon\right)\right)^2\right].
\label{equal45}
\end{align}
{Then, substituting (\ref{equal45}) into (\ref{equal44}), the upper bound for $R_D$ is obtained as shown in (\ref{equal43}).}

\section{Proof of Corollary 7}

By substituting (\ref{equal43}) into (\ref{equal37}), the approximate ESC can be represented as (\ref{equal114}),
where $t_0=\frac{2\pi\lambda_e}{\mu\left(\varpi\right){\rm e}^\frac{2v\left(\varpi\right)}{\mu\left(\varpi\right)}\Xi^2}=\frac{1}{\mu\left(\varpi\right){\rm e}^\frac{2v\left(\varpi\right)}{\mu\left(\varpi\right)}}\pi\lambda_eNK\nu\beta_0\rho_e\left[1-\frac{\epsilon^2}{{\frac{\pi}{4}\left(\epsilon+1\right)\left(L_\frac{1}{2}\left(-\epsilon\right)\right)}^2}\right]$, and (k) comes from the fact that when $\rho_d,\rho_e\rightarrow\infty$, the formula $\lim_{t_0\Gamma\left(t_1\right)\to\infty}{\left\{{\rm Shi}\left(t_0\Gamma\left(t_1\right)\right)-{\rm Chi}\left(t_0\Gamma\left(t_1\right)\right)\right\}}=0$ holds. We complete the proof by substituting the expressions of $\rho_d$, $\rho_e$, and $\mu_D$ into (\ref{equal114}).

\end{appendices}

\end{document}